\documentclass[graybox]{svmult}

\usepackage{type1cm}        % activate if the above 3 fonts are
\usepackage{makeidx}         % allows index generation
\usepackage{graphicx}        % standard LaTeX graphics tool
\usepackage{multicol}        % used for the two-column index
\usepackage[bottom]{footmisc}% places footnotes at page bottom

\usepackage{newtxtext}       % 
\usepackage[varvw]{newtxmath}       % selects Times Roman as basic font

\makeindex             % used for the subject index
\begin{document}

\title*{MAXI : Monitor of All-sky X-ray Image}
\author{Tatehiro Mihara, Hiroshi Tsunemi, Hitoshi Negoro}
\institute{
Tatehiro Mihara \at RIKEN, Wako, Saitama, Japan 351-0198 \email{tmihara@riken.jp} 
\and 
Hiroshi Tsunemi \at Osaka University,
Toyonaka, Osaka Japan 560-0043 \email{tsunemi@ess.sci.osaka-u.ac.jp}
\and
Hitoshi Negoro \at Nihon University, Tokyo, Japan 101-8308 \email{negoro.hitoshi@nihon-u.ac.jp}
}
\maketitle

\abstract*{Each chapter should be preceded by an abstract (no more than 200 words) that summarizes the content. The abstract will appear \textit{online} at \url{www.SpringerLink.com} and be available with unrestricted access. This allows unregistered users to read the abstract as a teaser for the complete chapter.
Please use the 'starred' version of the \texttt{abstract} command for typesetting the text of the online abstracts (cf. source file of this chapter template \texttt{abstract}) and include them with the source files of your manuscript. Use the plain \texttt{abstract} command if the abstract is also to appear in the printed version of the book.}

\section{MAXI mission} 
\label{sec:1}

%\section{MAXI mission}
%\label{sec:1}

Monitor of All-sky X-ray Image (MAXI) is a Japanese X-ray all-sky monitor onboard the International Space Station (ISS). 
We describe MAXI mission by adopting explanation from \cite{matsuoka2009}.

%%%%%%%%%%%%%%%%%%%%%%%%%%%%%%%%%%%%%%%%
% FIGURE 1: MAXI overview
%%%%%%%%%%%%%%%%%%%%%%%%%%%%%%%%%%%%%%%%

\begin{figure}[t]
\begin{center}
  \includegraphics[width=3.0in]{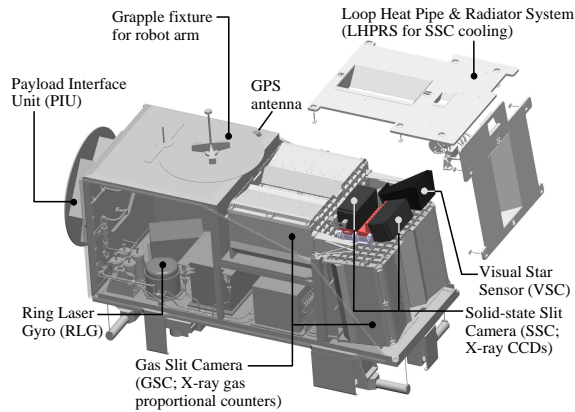}
\vspace{20mm}
\caption{MAXI overview. MAXI total weight: 520 kg.}  \label{MAXIoverview}
\end{center}
\end{figure}

The overview of MAXI is shown in Figure 1. 
 MAXI science instruments consist of two types of  
X-ray cameras, the Gas Slit Camera (GSC: \cite{mihara2011} \cite{sugizaki2011}) 
) and the Solid-state 
Slit Camera (SSC: \cite{tsunemi2010} \cite{tomida2011}).
Both GSC
and SSC employ slit and collimator optics.  
Their characteristics are listed in Table 1.  The support instruments 
consist of a Visual Star Camera (VSC) for a precise attitude determination, 
a Ring Laser Gyroscope (RLG) for a continuous tracking of the attitude, a
Global Positioning System (GPS) for an absolute timing and a Loop Heat Pipe and Radiation 
System (LHPRS).  The VSC  and the
RLG determine the directions of the GSC and the SSC as precisely 
as a few arc-minutes every second.
The LHPRS is used to cool down the
SSC camera body which is the hot side of  
the Peltier cooler of the CCD chip.

The MAXI payload was launched by the space shuttle Endevour on 2009 July
16, then mounted on the port No.~1 on JEM-EF.
MAXI started
nominal observation since 2009 August 15.  
The data are publically available from http://maxi.riken.jp .

ISS rotates by itself synchronously with its orbital motion 
so that one side always 
faces towards the center of the Earth.
Therefore, the sky side of JEM-EF has no Earth occultation of the field of view and surveys a great circle 
every ISS orbit.
MAXI has two Field of Views (FOVs). 
One is towards the moving-direction (horizon),
and the other is towards zenith.
The horizontal GSC cameras face a little upwards away from the horizon by 6 degrees
to avoid the Earth rim for any possible ISS attitude.
For SSC, the upward angle is 20 degrees 
to avoid ionized oxygen line from the upper atmosphere.
Although a considerably wide FOV is available, 
the ISS structure, moving solar paddles, docked space-shuttle and Space X vehicle partially block the view of JEM-EF.

%%%%%%%%%%%%%%%%%%%%%%%%%%%%%%%%%%%%%%%%
% TABLE 1
%%%%%%%%%%%%%%%%%%%%%%%%%%%%%%%%%%%%%%%%
\begin{table}
\begin{center}
\caption{Specification of MAXI slit cameras}\label{tab:camera}
\begin{tabular}{lcc}
\hline
  & GSC$^\dagger$: Gas Slit Camera &
    SSC$^\dagger$: Solid-state Slit Camera \\
\hline
\\
X-ray detector
		& {\parbox{30mm}{
		 12 pieces of one-\\
		 dimensional PSPC;\\
		  Xe 99 \% + CO$_2$ 1 \%,\\
		  1.4 atm at 0$^\circ$C}}
	    & {\parbox{30mm}{
			 32 chips of X-ray CCD;\\ 
			 1 square inch, \\
			 1024x1024 pixels}}
\\
X-ray energy range
		& 2--20 keV
			& 0.7--7 keV
\\
Total detection area
		& 5350 cm$^2$
			& 200 cm$^2$
\\
Energy resolution
		& 18 \% (5.9 keV)
			& $\leq$ 150 eV (5.9 keV)
\\
Field of view$^*$  
		& 1.5 x 160 degrees
			& 1.5 x 90 degrees
\\
Slit area for camera unit$^\dagger$ 
		& 12.3 cm$^2$
			& 1.35 cm$^2$
\\
Detector position resolution 
		& 1 mm $^\ddagger$
			& 0.025 mm (pixel size)
\\
Localization accuracy 
		& 0.2 deg
			& 0.2 deg
\\
Absolute time resolution
		& 50 $\mu$s
			& 5.8 s
\\
Weight
		& 160 kg 
			& 11 kg
\\
\hline
\end{tabular}
\end{center}
Notes.  
\par\noindent
$*$  FWHM $\times$ Full-FOV.
\par\noindent
$\dagger$ GSC consists of 6 camera units. Each unit
consists of two PCs.  SSC consists of two camera units, SSC-Z and SSC-H.
Effective area for one source is typically 6 cm$^2$ for one GSC camera when X-ray enters vertically to the detector. 1 cell width = 32 mm, with 3 beams shadows (1.4 mm width), and slit width 3.7 mm makes 6.1 cm$^2$. 
\par\noindent
$\ddagger$ For 8 keV at 1650 V. Most photons are softer than this, 
and many detectors were operated at 1550V to avoid spark.
Observed position accuracy is the sum of this and slit width.
It becomes 1.5 degrees or a little worse.
\end{table}
%%%%%%%%%%%%%%%%%%%%%%%%%%%%%%%%%%%%%%%%

The MAXI proposal was finally accepted in 1997 by the National Development
Space Agency of Japan (NASDA), now known as the Japan
Aerospace Exploration Agency (JAXA).  MAXI is the first 
astronomical payload for JEM-EF on the ISS. Although the launch was 
scheduled for 2003, the space shuttle, Columbia accident, and the ISS 
construction delay resulted in the postponement of the MAXI launch  
to 2009.

 A specific object is observed by MAXI with a slit camera for a 
limited time with every ISS orbit. 
The X-ray detector of the camera is sensitive to a one 
dimensional image through the slit, where the rectangular wide FOV 
through the slit spans the sky perpendicular to the ISS moving 
direction as shown in Figure 2 in case of GSC 
($ \pm 1.5^\circ \times \pm 40^\circ $).  The scanning image of an 
 object is obtained 
with the triangular response of a slat collimator according to ISS 
rotation. Combination of the slit image and 
the triangular transmission image makes 
a source image in the sky, called as a Point Spread 
Function (PSF) (Figure 3). One-side of PSF is determined by the slats-collimator 
(1.5$^{\rm o}$ for all energies), while the other side of PSF is determined 
by the position detection capability of the proportional counter 
(large in low energies, and small for high energies with a tail for a slant incident angle in high energies). 

\begin{figure}
  \includegraphics[width=2.5in]{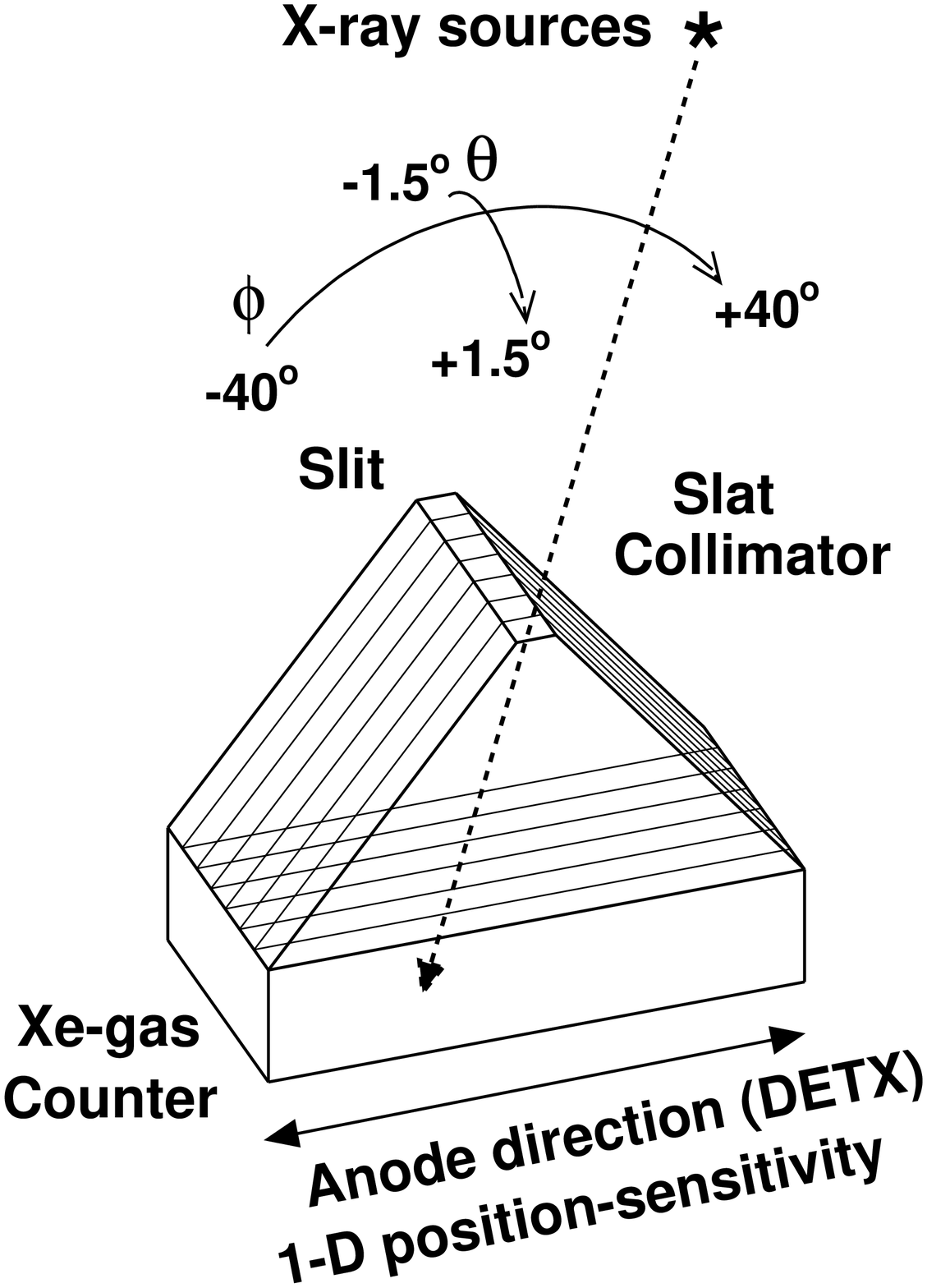}
  \includegraphics[width=2.5in]{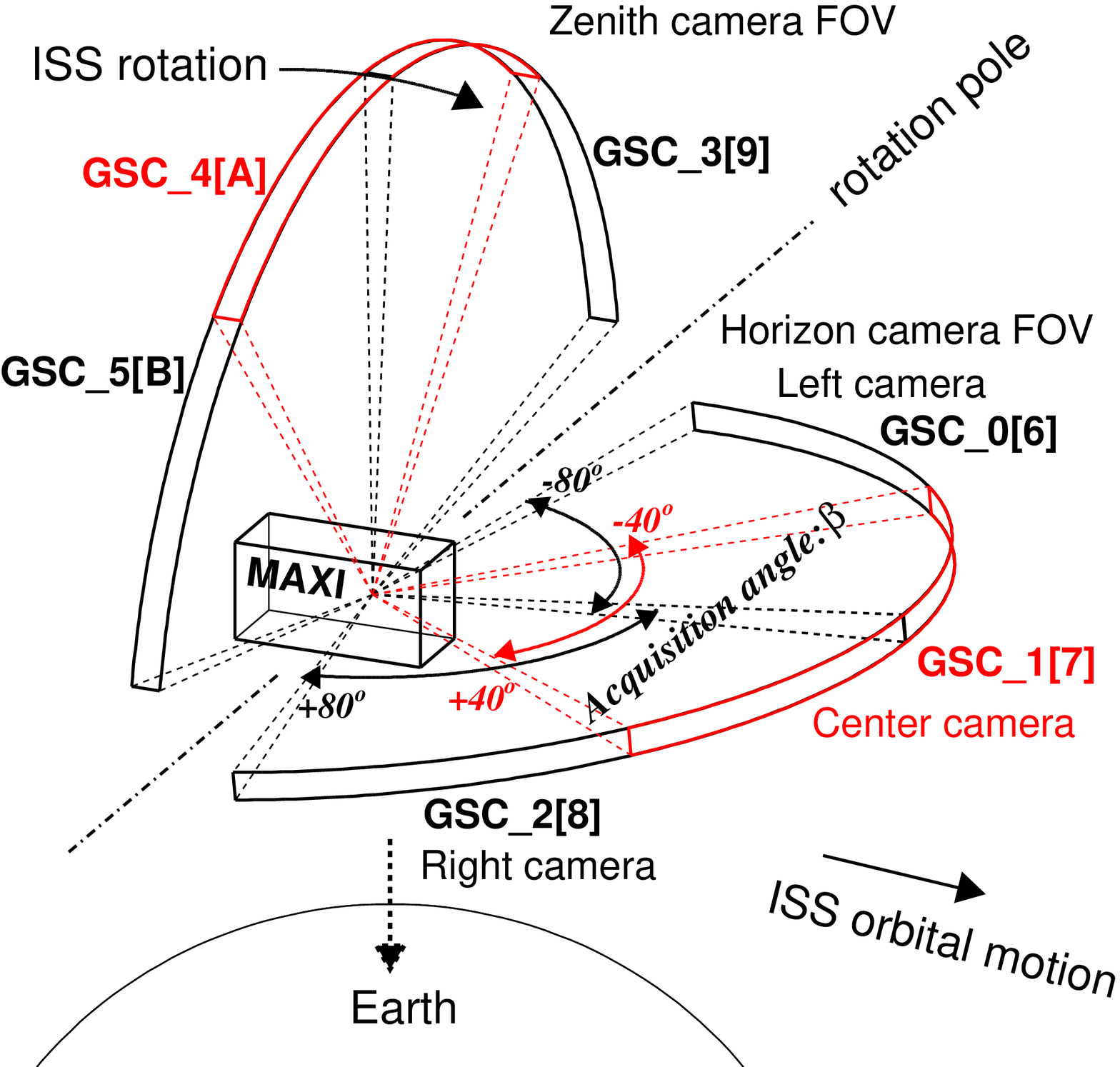}
  \vspace*{0mm}
\caption{Principle of MAXI Slit cameras. It consists of slit 
\& slat collimators and one-dimensional position sensitive 
X-ray detector.}  \label{slitcamera}
\end{figure}

\begin{figure}
\sidecaption
    \includegraphics[width=65mm]{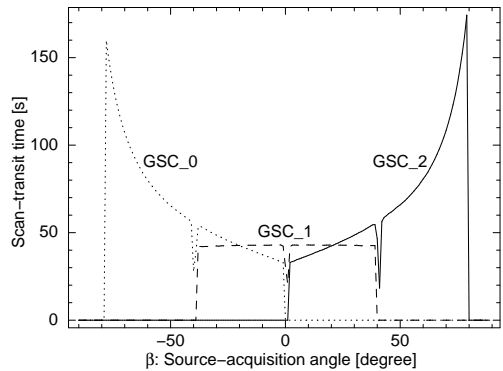} 
    \caption{ Scan-transit time as a function of source-acquisition
      angle for each GSC unit of GSC\_0, GSC\_1, and GSC\_2, in the
      ISS normal attitude. }
    \label{fig:transitduration}
 \end{figure}

\label{sec:scan_transit}
The transit time of a point source is
40--150 seconds with the FOV of 1.5$^\circ$-width (FWHM) every
92-minute orbital period.  
Figure
\ref{fig:transitduration} illustrates the relation between the
scan-transit time on each GSC counter and the source acquisition angle $\beta$
from the rotation equator in the ISS normal attitude (see the geometry in
figure \ref{slitcamera}). The scan
duration increases from the ISS rotation equator ($\beta=0^\circ$)
towards the pole ($\beta=\pm 90^\circ$).
Objects located along a great circle stay for 45 s in the 
FOV of MAXI cameras, where the time of stay is the shortest in the 
MAXI normal direction. For  objects in the slanted 
field away from the great 
circle, the observation time becomes slightly longer.  

 \begin{figure}
\sidecaption
 \includegraphics[width=60mm]{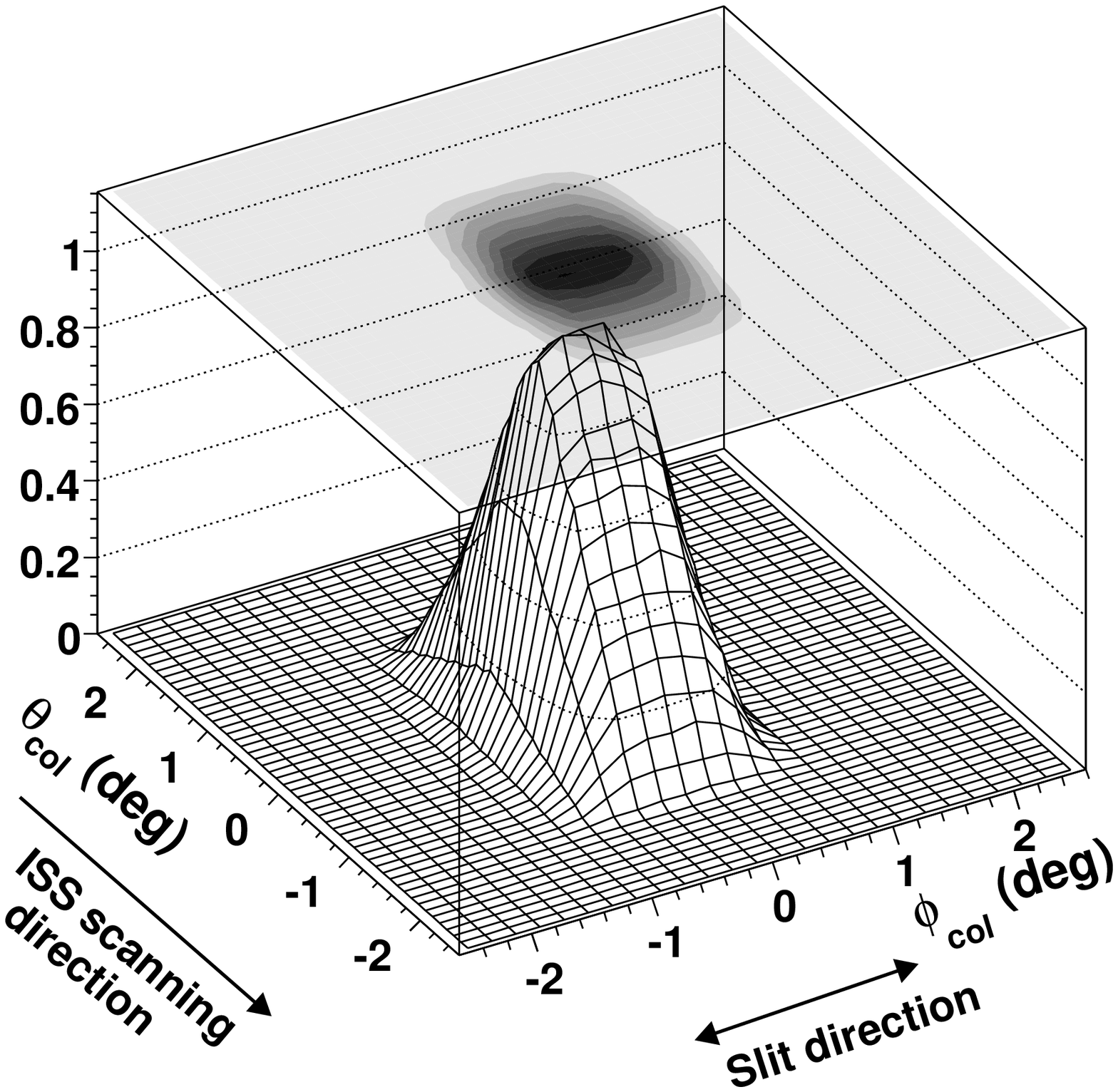}
\caption{Typical sample of PSF (Point Spread Function); a slat 
collimator has a triangular response (ISS scanning direction), 
while a slit and X-ray detector make a trapezoidal response.
By monte-carlo simulation for incident angle 10 degrees and X=ray energy 5 keV for GSC.}\label{fig:psf}
\end{figure}

\begin{figure}
\sidecaption
\includegraphics[width=70mm]{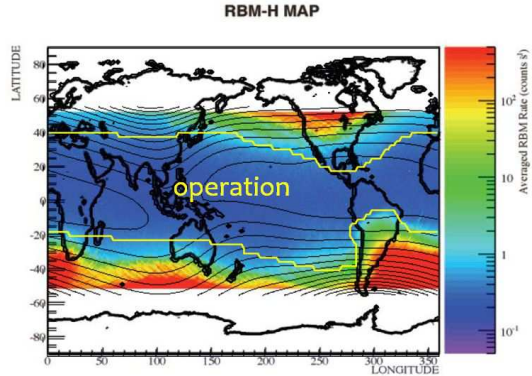} 
\caption{ Count rates of Radiation belt monitor - Horizon (RBM-H) onboard MAXI.
GSC is operated in the low background region, which is shown between the yellow lines. 
The MAP is set to MAXI and on/off of high voltage of GSC is automatically 
controlled.}
\label{fig:rbmmap}
\end{figure}

Any target will come 
repeatedly in each field of the horizontal and zenithal cameras with 
every ISS orbital period.
So in principle, a source is scanned twice in one orbit.
But in fact, since the operation region is limited by the radiation zone,
it is usually scanned only once in one orbit (Figure \ref{fig:rbmmap}).
As a matter of fact, MAXI coverage of the sky is as 80\% in 1-orbit (92minutes) and 89\% in 1-day.

The scanning direction and camera for a source changes with the 72 days cycle of 
ISS orbital precession. 
The sky near the scan poles comes out to the observable 
region in about two weeks. GSC is off when a part of the FOV of the camera comes at 5 degrees from  the sun.
However, the sum moves on the sky by 10 degrees in 10 days.
Therefore there is no permanent invisible part of the sky.
Because of these reasons the MAXI light curves 
often suffer a spurious 72-day oscillation.

MAXI data will be down linked through the Low Rate Data Link (MIL1553B), 
and the Medium Rate Data Link (Ethernet).  
The basic data is contained in both rate data. 
MAXI is operated through the 
Operation Control System (OCS) at Tsukuba Space Center (TKSC), JAXA.  
MAXI data are processed and analyzed at TKSC, and also they are 
transferred to the Institute of Physical and Chemical Research
(RIKEN).  General users of MAXI can request the 
scientific data from RIKEN.

This nova alert system has been developed  to achieve
automatic alert in section 4 (\cite{negoro2016}). 
The MAXI team will also maintain archival data at RIKEN using 
the data from 
low-rate data and medium-rate data, such as all sky X-ray images, 
X-ray light curves of dedicated X-ray sources. 
In principle all astronomical data of MAXI will be available
for public distribution (\cite{kohama2009}).
MAXI on-demand can process any part of the sky and any time in the mission
(\cite{nakahira2013}).

%1. MAXI mission  2p
%@@slit camera, scan, detectors, DP, cooling, power,tÌâp

\section{GSC}
%1.1  GSC  3p

% GSC section

The GSC is the main instrument of MAXI.
We describe GSC by adopting explanation from \cite{mihara2011}.

The MAXI/GSC employs slit camera optics.  The slit camera has an
advantage of being free from the source contamination over the
coded-aperture mask while it has a disadvantage in the limited slit
area. To achieve the high sensitivity, large-area
proportional counters filled with Xe gas are used for the X-ray
detectors.  The total detector area of 5350 cm$^2$ using twelve gas
counters is optimized within the limit of the payload size
(0.8$\times$1.2$\times$1.8 m$^3$).

Each set of six cameras is assembled into a module with the same FOV. One module points towards the direction of the ISS motion and the other towards the zenith. 
They are
named as horizon and zenith modules, respectively.  Each horizon/zenith
module covers a wide rectangular FOV of 160$^\circ$$\times$1.5$^\circ$
(FWHM) with an almost equal geometrical area of 10 cm$^2$.  
The 10$^\circ$ angle at the
FOV edges on both the rotation poles are not covered because they are  
obstructed by the ISS structures.

The two FOVs of the horizon and the zenith modules 
would scan almost the
entire sky in the 92 minutes orbital period if there were no high background region.
In this case, any X-ray source is to be
observed twice in one orbit.
The horizon module scans first then followed by the zenith module 
after 21.5 minutes. The separation of the two FOV is 84$^\circ$.  
Both FOVs have no Earth
occultation and do not use any moving mechanics.

In the actual in-orbit operation, observation periods are limited in a
low particle-background area in order to protect the counters from the
heavy particle irradiation.  It reduces the observation duty cycle 
down to $\sim$40\% \cite{sugizaki2011}.
The two FOVs are capable of covering the whole sky, although the scan frequency becomes once in an orbit.

\subsection{Gas counter}

The GSC proportional counters employ resistive carbon fibers with a
diameter of 10 $\mu$m for anode wires.  Higher resistive anodes
are preferable for the better position resolution because the thermal
noise on the readout signal is inversely proportional to the anode
resistance.  A carbon fiber is better than a nichrome wire.
Although the traditional carbon-coated quartz has a larger
resistance, it is mechanically weak, thus easy to break by launch vibration.
The carbon fiber anodes in the
GSC were 
developed at RIKEN and were successfully used in
the HETE/WXM \cite{shirasaki2003}.  

\begin{figure}
%  \begin{center}
  \includegraphics[width=60mm]{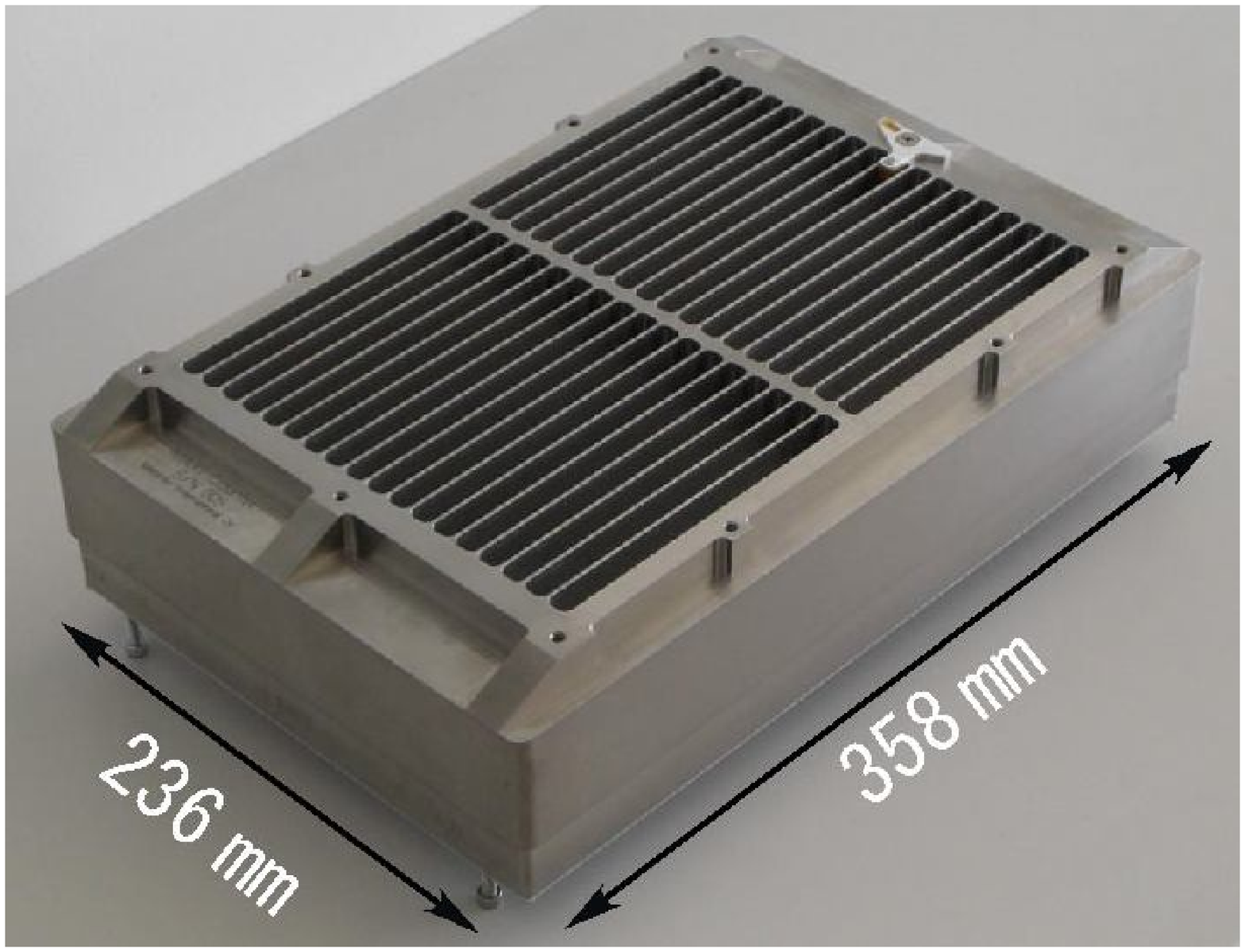} 
%    \FigureFile(85mm,){fm005a.ps} 
    \hspace{5mm}
      \includegraphics[width=50mm]{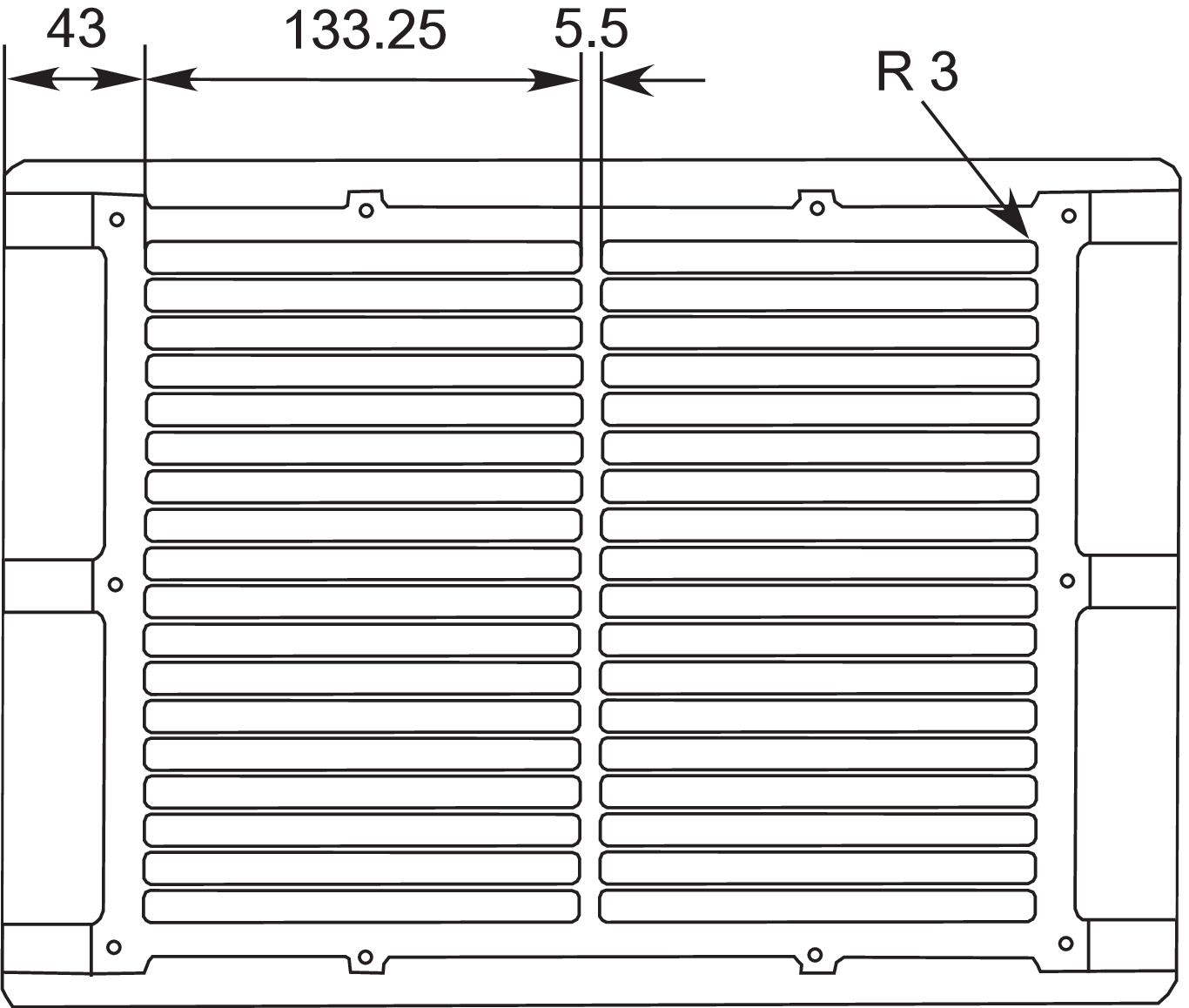} 
%    \FigureFile(70mm,){window.eps} 
  \caption{ Proportional counter used in GSC
    and drawing of the window from the top.}
  \label{fig:counter}
%  \end{center}
\end{figure}

Figure \ref{fig:counter} shows a picture of a single GSC proportional
counter.  All the flight counters were manufactured by Metorex (now a
part of Oxford Instruments) in Finland.  The front X-ray window has an
area of 192 $\times$ 272 mm$^2$. It is sealed with a 100-$\mu$m-thick
beryllium foil.  To support the pressure on the beryllium foil in
vacuum that amounts 7300 N in the whole area, grid structures with a
17-mm height are placed every 10.6-mm pitch parallel to the anode
wires.  The vertical grid is placed only at the center to keep the
open area as large as possible.  The maximum pressure of 1.66
atm is expected at the temperature of 50$^\circ$C.  
Every flight counter was tested to
withstand 1.5 times higher than the design pressure, i.e.\ 2.5 atm.
The bodies of the gas counters were made of titanium, 
which has sufficient strength and a heat expansion coefficient 
close enough to that of beryllium.
The beryllium foil is glued on the body with epoxy.
%
%
%Another difference from the WXM PC is the shape of the cells.  The
%cross section of the cells of the GSC PC is more rectangular to
%decrease the number of signals from eight square cells to six
%rectangular ones.  The bottom and side vetos were placed surrounding
%the carbon-anode cells.  

\begin{figure*}
  \begin{center}
    \includegraphics[width=120mm]{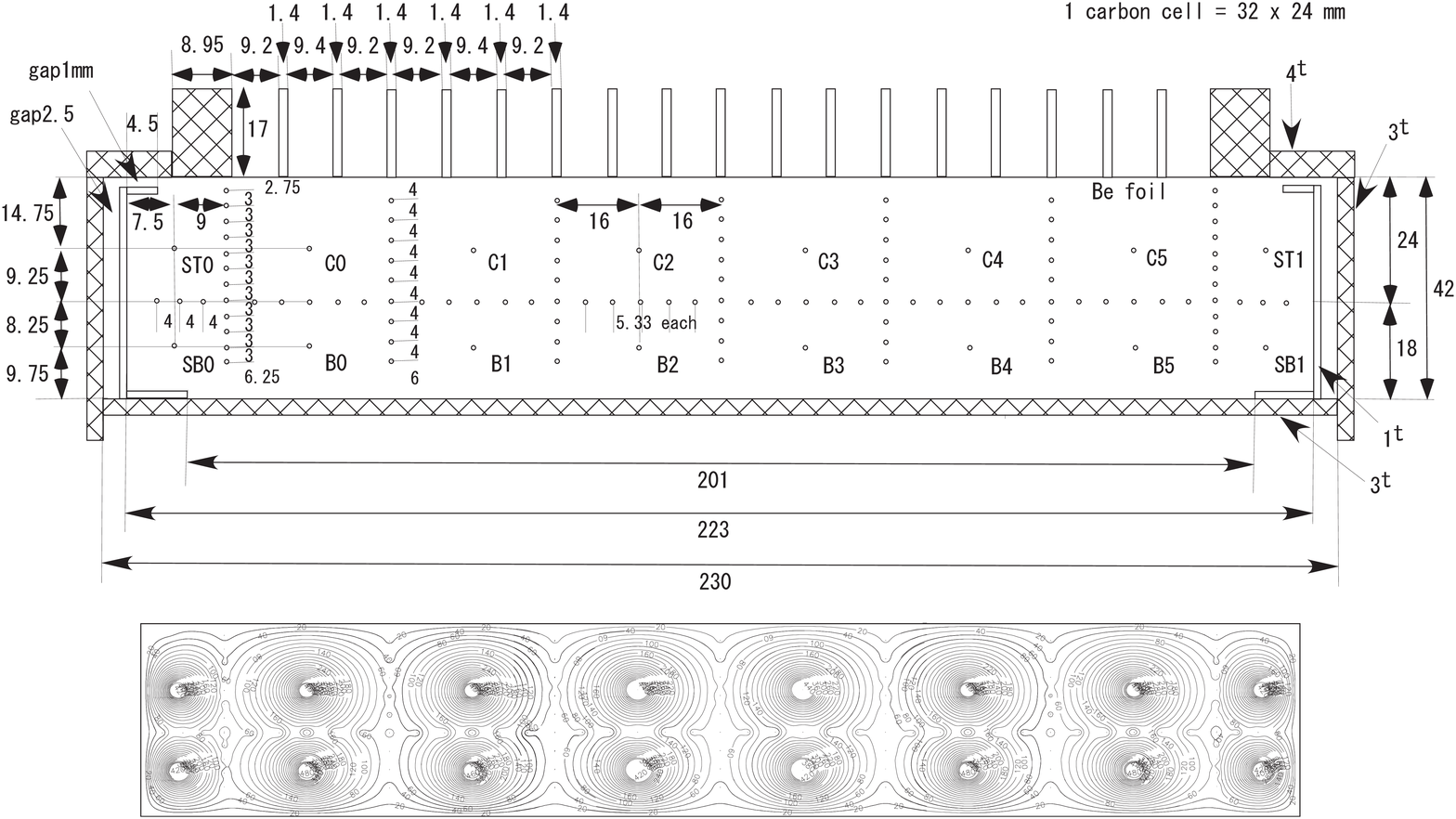} 
%    \FigureFile(150mm,){locationVdenba.eps} 
%    \FigureFile(85mm,38mm){locationofwiresbeamronbunVdenba.epsf} 
%    \FigureFile(70mm,12mm){Vdenba.ps} 
% canvas epsf de save, icon nasi.
  \end{center}
  \caption{ Cross-section view of GSC proportional counter 
    \textcolor{black}{
    on the plane perpendicular to the anode wires.}  All
    anode/ground-wire locations are shown. The names for anode, C, B,
    ST and SB, denote Carbon, Bottom, Side Top and Side Bottom,
    respectively.  The wires of B0 to B5 are connected together in the
    counter and read out as a single bottom-veto (BV) signal.  The same for ST0,
    ST1, SB0 and SB1, as a side-veto (SV) signal. 
    All numbers represent the scale
    in units of mm. Electric potential in the counter calculated by
    Garfield is shown in the below. The ``Garfield'' is a program to
    simulate gas counters developed in CERN
    (http://garfield.web.cern.ch/garfield/).}
  \label{fig:wiregrids}
\end{figure*}

Figure \ref{fig:wiregrids} 
shows the counter cross-section views.  
The gas cell is divided by ground wires into six carbon-anode cells 
for X-ray detection and ten tungsten-anode cells for particle veto.
The carbon-anode layer and the bottom veto-detector layer have depths of
24 mm and 18 mm, respectively.  These sizes are determined so that the
main X-ray detectors and the veto detectors have enough efficiencies
for X-rays in the 2--30 keV band and minimum-ionization particles, 
respectively. The minimum-ionization energy in the 18-mm thick Xe
gas is 30 keV.  

The carbon anodes and veto anodes are not located at the center of each cell in the vertical direction.
The anode locations, 
the aspect ratios of these gas cells, and the 
spacings of the ground wires are determined so that the spatial
non-uniformity of gas gain is small within each cell.  

\textcolor{black}{
The tension of the carbon-anode wire
is set to 4 gW, which is 
sufficiently smaller than the breakage limit,
$\sim25$ gW.
All anode and ground wires are fixed via a spring at right end to 
absorb the difference of the heat expansion coefficient and keep the wires
tight and straight.}
The veto anode wires are
made of gold-coated tungsten with a 17-$\mu$m diameter,
\textcolor{black}{
which is pulled with a tension of 18 gW. We chose
as thin wires as possible for veto anodes
to achieve similarly high gas gain as the carbon anodes }
since the same high voltage (HV) is applied to both the carbon anodes and the
veto anodes. The gas-gain ratio of carbon anode to the
veto anode is 20:1.  The ground wires are made of gold-coated tungsten
with a 50 $\mu$m diameter.
\textcolor{black}{
Each tension is about 50 gW.
}

We tested several kinds of gas mixture and chose a combination of
Xe (99\%) + CO$_2$ (1\%) with a pressure of 1.4 atm at 0$^\circ$C.  
The amount of CO$_2$ is decreased
from WXM PC (3\%) \cite{shirasaki2003} in order to reduce the spatial
gas-gain non-uniformity, and still keep the sufficient
quenching effect \cite{mihara2002}.

The position resolution and the energy resolution are 
incompatible requirements.
The position resolution is primarily determined by
the thermal noise on the resistive-anode wire against the readout
signal charge.  The higher gas gain is basically preferred for the
better position resolution.  However, the high voltage for the best
position resolution is usually in the limited proportionality region
rather than the proportionality range, where the 
spatial gain non-uniformity is larger due to the space-charge effect, 
which also degrades the energy
resolution.  The operating high voltage (HV = 1650 V and the gas gain
of 7,000) 
is chosen to achieve a sufficient position resolution
and still keep an adequate energy resolution.

%As the resistance of the carbon
%fiber is lower, Johnson noise is higher.  

%In order to obtain better position resolution, we have to operate gas
%counter in rather high voltage .
%Then the anomalous gas-gain (gain non-uniformity in the cell) appears and
%makes the energy resolution worse.

%The phenomena is that
%the gain depends on the X-ray stopped position in the cell.

For the in-orbit calibration, 
a weak radioactive isotope of $^{55}$Fe is installed 
in every counter, which 
illuminates a small spot of about 1 mm in diameter
at the right end of the C2-anode cell 
(almost middle of the upper side in the left photo in Figure
\ref{fig:counter})
Each isotope has a radiation of 30 kBq
and its count rate by a GSC counter is about 0.2 c s$^{-1}$ at the
launch time.

\subsection{Electronics}

Each GSC counter has 
six position-sensitive anodes readout at the both ends
(left and right), and two signals for connected veto anodes.
A total 14 preamplifiers are used for the 14 analog signals.
We selected a hybrid-IC, Amptek A225, for the preamplifier, which is
made with a space-use quality and has a low-power consumption.
The feedback capacities of veto anodes are
left as they are 1 pF at the default, while those of carbon anodes are
modified to 4 pF by adding an external 3 pF capacitor
in order to obtain closer pulse heights for both signals from carbon anodes and veto anodes. The ratio becomes 5:1.

The maximum count rate in the current operation mode of MAXI is 40 Crab for a persistent source (lasting for the full scan length of 45 s), 
which is limited by the amount of event buffer and the 64 bit length of an event.
Meanwhile, the maximum rate for the shorter time scale than 20 s is $\sim$100 Crab,
which is determined by the event transmission rate of the electronics.

%---------------------------------sugizaki
\subsection{Background in orbit}

\begin{figure}
  \begin{center}
    %\FigureFile(8.5cm,){bgdspec_sens3_ph.eps} 
    \includegraphics[width=8.5cm]{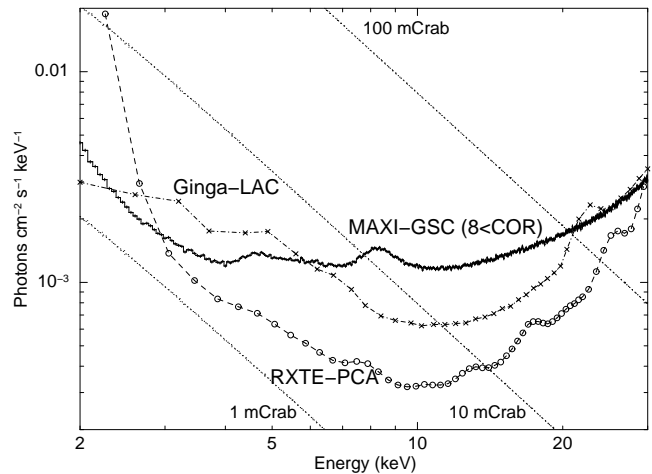} 
  \end{center}
    \caption{GSC background energy spectrum normalized  by effective area
      and comparison with those of Ginga-LAC, RXTE-PCA, and
      Crab-like source spectra.  Expected contributions of the
      CXB in each instrument are included.  }
    \label{fig:bgd_spec_uf}
\end{figure}

Figure \ref{fig:bgd_spec_uf} illustrates the background spectrum
normalized by effective area, and those of the Ginga LAC
(\cite{hayashida1989}) and the RXTE PCA (\cite{jahoda2006}),
where expected contribution of the CXB in each instrument are
included.  Crab-like source spectra are shown together as the
comparison.  The GSC background is approximately 2 mCrab at 4 keV and
10 mCrab at 10 keV.  The level is almost comparable to that of the
Ginga LAC and slightly higher than that of the RXTE PCA.

\begin{figure}
  \begin{center}
    %\FigureFile(8.5cm,){detxprof0345_pi40_200_20090817.eps}
    \includegraphics[width=8.5cm]{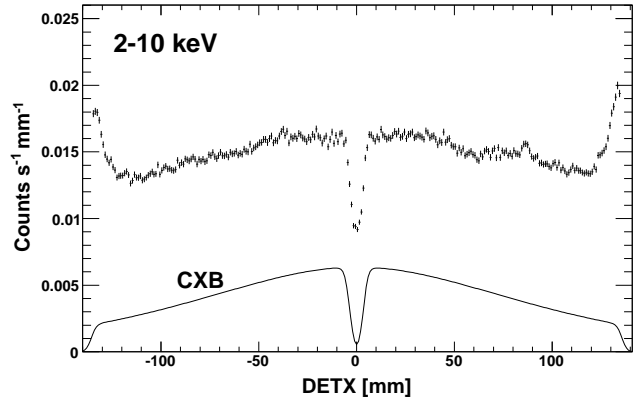}

  \end{center}
    \caption{ Background spatial distribution in GSC gas counters
      along anode wires (DETX) in each energy band of 2-10 keV.
      Estimated contributions of the CXBs are
      shown with solid lines.  The dips at the center (DETX$=$0) are
      due to the shadow of the support structure of the counter
      beryllium window.  }
    \label{fig:bgd_detx}
\end{figure}

Figure \ref{fig:bgd_detx} shows spatial distributions of the residual
backgrounds along the anode wire in 2-10 and 10-30 keV energy bands
and the estimated contributions of the CXB.  Each GSC counter has a
sensitivity gap at the center of the anode wire due to the frame
structure to support 0.1-mm thick beryllium window. It causes the dip
at the center in the spatial distribution.  The depth of the dips can
be mostly explained by the extinction of the CXB component at the
support structure.

The profile of the residual background left after subtracting the CXB
component is primarily flat and slightly increases at both ends of the
anode wires.  It is because the efficiency of the anti-coincidence
reduction decreases at the counter edge.  The sensitivity is expected
to get worse in these area.

Figure 2 in 
\cite{mihara2014} is the background count rates in 4-10 keV band. The background rates 
sometimes went down to a half for some tens of days. 
These spans are when Soyuz spacecraft was not docked at the closest docking port.
The gamma-ray altimeter of Soyuz spacecraft has a strong isotope (~0.8 Ci of $^{137}$Cs and $^{60}$Co). It doubles the 
background of MAXI/GSC.

\begin{figure}
  \begin{center}

    \includegraphics[width=12cm]{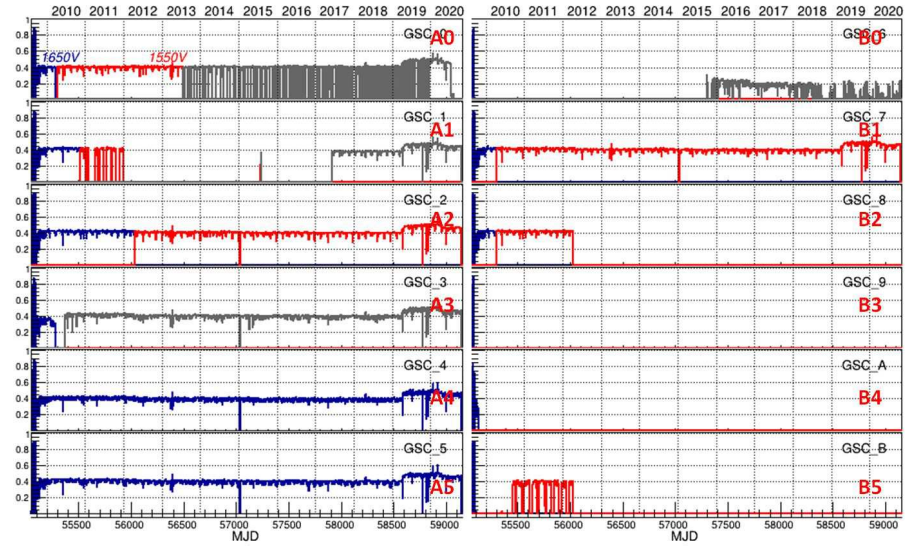} 

  \end{center}
    \caption{ Daily HV-on duty cycle of 12 GSC cameras from 2009 to 2020. Period of the standard operation with 1650 V is in blue, and 1550 V in and red lines. Those of the special condition is in gray lines.
    }
    \label{fig:worktime}

 \end{figure}

Figure \ref{fig:worktime} is the duty cycle (functioning time) of GSC. 
The average duty cycle of 40\% comes from the limited operation on orbit  where backgrounds by charged particles are low.  
At the time of 2020, two counters, camera 3, camera 6 are operated under the limited functionality (malfunctional veto counter). 

On 2013 June 15, the gas gain of GSC camera 0 suddenly started to rise (\cite{mihara2014} Figure 4). A small micrometeoride with about 50 
$\mu$m in diameter hit and cracked the Be window to   
cause a small gas leak. 
Since the gas leaked out, the operation of GSC camera 0 was stopped on 2020 August 15.

%2.1 GSC All-sky maxp and catalog 1p@@O´

\section{SSC}
%1.2. SSC 3p
%%\title*{Contribution Title}
%%\author{Name of First Author and Name of Second Author}
%%\institute{Name of First Author \at Name, Address of Institute, \email{name@email.address}
%%\and Hiroshi Tsunemi \at Department of Earth and Space Science, Graduate School of Science, Osaka University,
%%1-1 Machikaneyama, Toyonaka, Osaka 560-0043 %%\email{tsunemi@ess.sci.osaka-u.ac.jp}}
%

%%\maketitle

\subsection{X-ray CCD and its function}

\begin{figure}
   \includegraphics[width=2.5in]{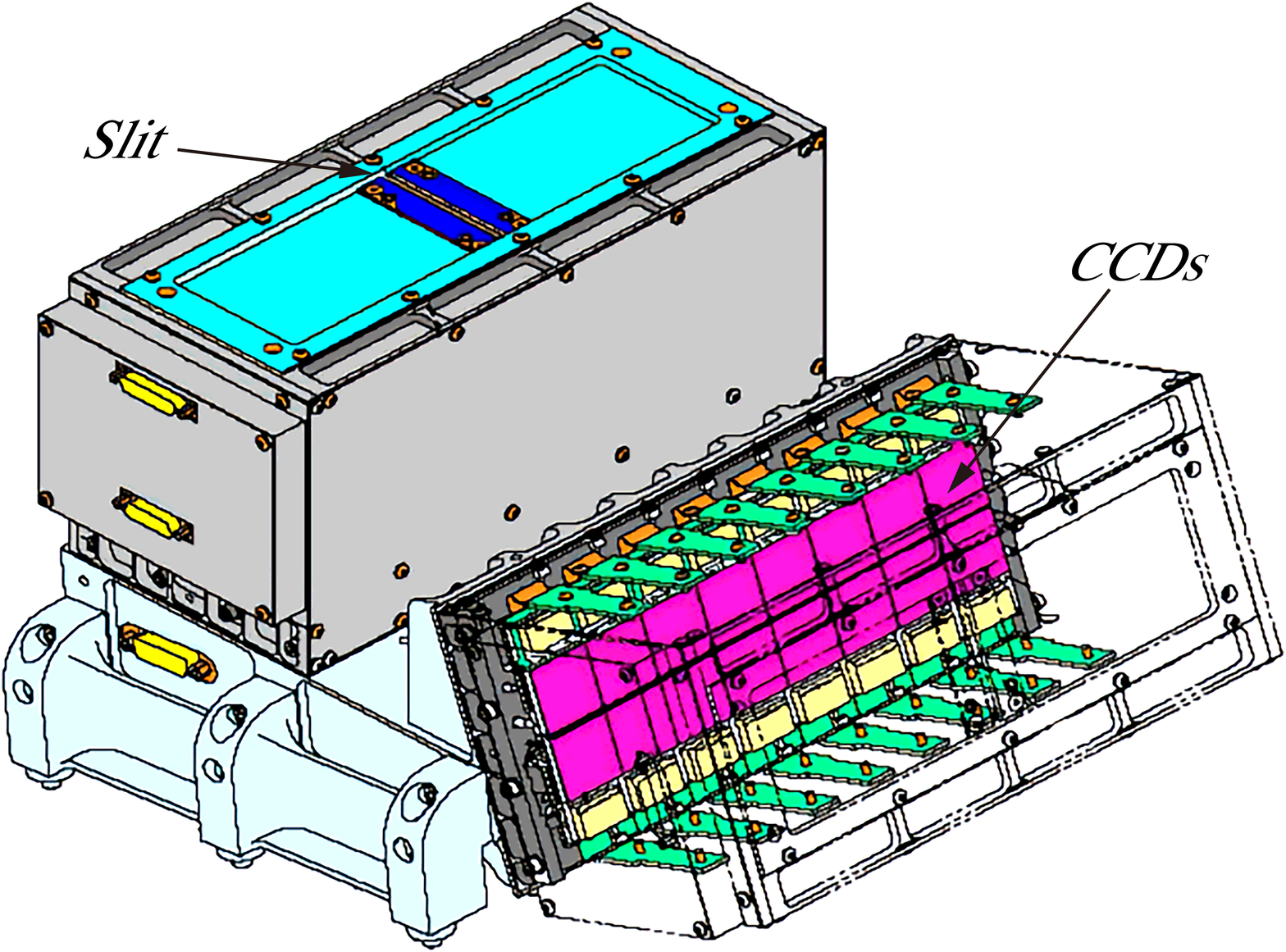}
   \includegraphics[width=2.5in, bb=0 0 2304 1728]{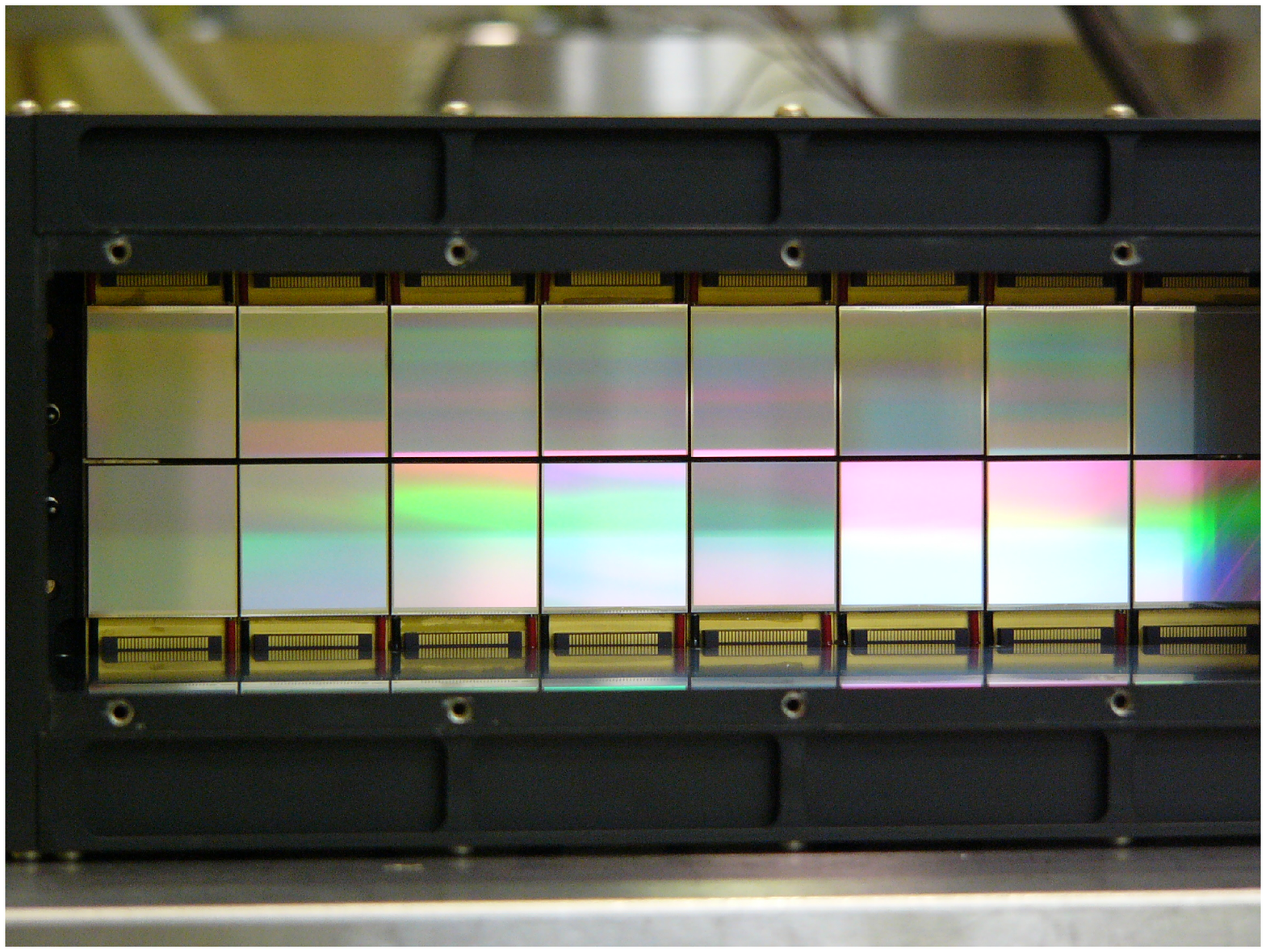}
\caption{Left is an exploded view of the SSC.  Two SSCUs are set on the SSC stand.  Right is a photograph of CCD array in an SSCU.  The gap width between the adjacent CCDs is 0.4\,mm.}  \label{SSC_plain}
\end{figure}

The SSC \cite{tsunemi2010} consists of two identical SSC Units (SSCUs).  They are thermally connected to an aluminum stand as shown in Fig.\,\ref{SSC_plain}.  Each SSCU consists of CCD units, preamplifiers, multiplexers, a collimator and slit unit, and a calibration source.  The ISS orbits in a fixed orientation to the Earth so that MAXI always sees the sky.  One SSCU (SSC-Z) monitors the zenithal direction and the other SSCU (SSC-H) monitors +20$^\circ$ above the horizontal direction.  Each has 16 CCDs in 2$\times$8 array, as shown in Fig.\,\ref{SSC_plain} \cite{tomida2011}.  Each CCD is a front-illuminated device, manufactured by {\it Hamamatsu Photonics} \cite{tsunemi2007}.  It is 25\,mm square with a 24\,$\mu$m square pixel ($1024\times 1024$).  Therefore, the geometrical area of the SSC (32 CCDs) is about 200\,cm$^2$.  An aluminum coating of 200\,nm thickness on the CCD enables it to eliminate an optical blocking filter in front of the CCD, which in turn makes it possible to eliminate a vacuum tight body.  The CCD is fabricated on an high-resistivity wafer, providing a thick depletion layer of about 70\,$\mu$m.  
\begin{figure}
\sidecaption
   \includegraphics[width=2in, bb=0 0 643 495]{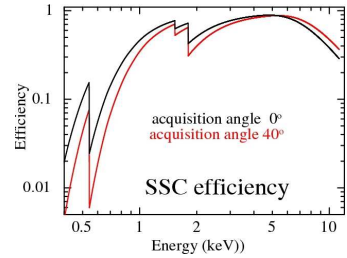}
\caption{Detection efficiency of the SSC calibrated by the Crab nebula spectrum.  We employ G0 for E$\le$1.8\,keV and G0+G1+G2 for E$\ge$1.8\,keV.}  \label{SSC_QE}
\end{figure}

The SSC has no X-ray mirror, therefore the energy range of the SSC is determined by mainly CCD wafer.  The quantum efficiency for soft X-ray is limited by absorption at the gate structure (dead layer) and aluminum coating on the front surface of the CCD wafer.  That of hard X-ray is limited by the thickness of the depletion layer.  With taking into account the electronics, the effective energy range of the SSC is 0.5-12\,keV.  The SSCU has an entrance slit of 2.7\,mm width just above the center of the CCD array that restricts the field of view (FOV) to 90$^\circ $.  Since the slit is just above the center of the camera, an acquisition angle of X-rays passing through the slit depends on the location of the CCD, resulting a different detection efficiency for each CCD.  Fig.\,\ref{SSC_QE} shows the detection efficiency of the CCD by calibrating the Crab nebula \cite{tsunemi2010}.  We placed collimator sheets between the slit and the CCD array that restricts the FOV to $1 \overset{\circ}{.} 5$.  Since the slit and the collimator sheets are perpendicular to each other, we obtain the fan-beam FOV of the SSC, 90$^\circ \times 1\overset{\circ}{.} 5$.  In this way, we run CCDs to function as one-dimensional imager.  One orbit of the ISS covers 90$^\circ \times 360^\circ$.  The sky coverage region gradually shifts according to the precession of the ISS orbit.  In this way, we can cover the entire sky about every 70-day.

The data handling scheme of the MAXI/SSC is similar to that of the previous satellites, which is given in the literature \cite{tomida2011}.  There is one video chain for each camera (16 CCDs) that runs CCDs in parallel sum mode where charges in multiple pixels are summed on-chip.  In the standard operation, we obtain 16$\times$1024 pixel data for each CCD.  Since we sequentially read 16 CCD chips, the read-out time is 5.865\,s that depends on the number of rows of the on-chip sum.

The inclination of the ISS orbit is $51 \overset{\circ}{.}6$, resulting to pass at higher latitude regions than
other X-ray missions in low-earth orbit.  We switch off the bias level to the CCD during the passage of the South Atlantic anomaly (SAA) \cite{miyata2003}.  However, we keep switching on during the passage of high latitude region where the background is unstable and becomes too high to obtain good data.  We employ event recognition method similar to the ASCA grade \cite{gendreau1995}.  In the parallel sum mode, the charge spread of the signal is effectively valid only for G0 (single event), G1 (left split event), G2 (right split event) and G3 (three-pixel event).  We do not see the charge spread in the vertical direction, therefore the background rejection efficiency is worse than that of the normal mode in other satellites.  We expect that the X-ray events form G0, G1 or G2 while the charged particle event forms G3.

Furthermore, it degrades a performance of the CCD to pass through high particle background regions.  The charged particles degrade the CCD performance by generating charge traps and by increasing dark current.  Therefore, the CCD employed has two characteristics for radiation hardness.  One is a notch structure that confines the charge transfer channel to a very narrow width.  This actually improves the radiation-hardness by a factor of three \cite{tsunemi2004}.  The other is a charge-injection (CI) through which we can continuously inject some amount of charge at every 64 rows in the 64 binning mode \cite{miyata2002}.  The CI method can partly compensate for any degradation of the charge transfer inefficiency (CTI) of the CCD \cite{tomida1997}.  This technique \cite{prigozhin2008} was, for the first time, added to the Suzaku XIS \cite{koyama2007} in orbit and improved the performance of the CCD \cite{uchiyama2009}.

\subsection{Cooling system}

\begin{figure}
\sidecaption
   \includegraphics[width=2.5in, bb=0 0 486 541]{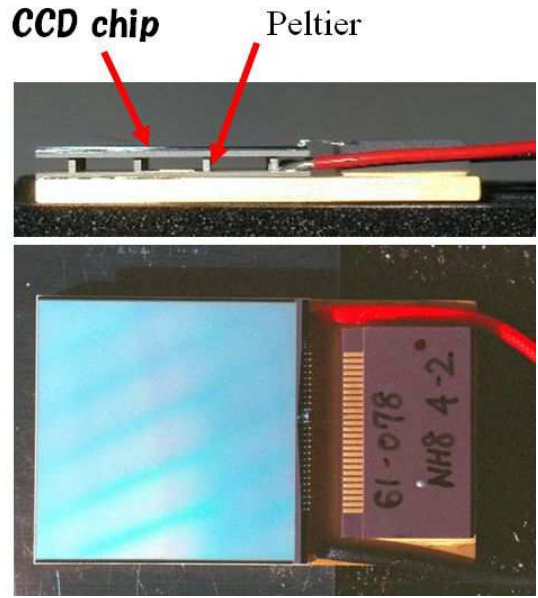}
\caption{Side view and top view of a CCD in the SSC.}  \label{SSC_chip}
\end{figure}

CCDs for X-ray detection in photon counting mode requires a low working temperature.  The cooling system of the MAXI/SSC consists of two parts: one is a peltier device and the other is a radiator through a loop heat pipe (LHP).  As shown in Fig.\,\ref{SSC_chip}, the CCD chip (wafer and its silicon base) is mechanically supported by a single-stage peltier devices (24 posts made of bismuth telluride) so that we can cool the CCD chip.  There are bonding wires between the CCD chip (cold part) and its surrounding hot part.  Two SSCUs are placed on the SSC stand that is thermally connected to two radiator panels.  One radiator panel (radiator-Z, 0.527m$^2$) is facing to the zenith and the other (radiator-H, 0.357m$^2$) is facing to the horizon.  They are designed such that we can obtain the maximum area within the allocated volume of the MAXI.  The radiator temperature depends on the day/night condition of the ISS.  The LHP works as a heat-diode so that the colder radiator is selected to receive the heat from the SSC stand.

\begin{figure}
\sidecaption
   \includegraphics[width=2.5in]{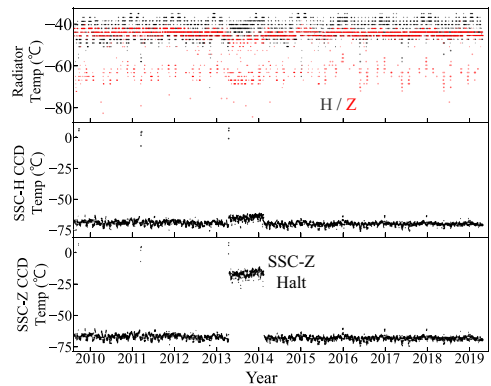}
\caption{Thermal history of the radiator-Z/H, CCDs in SSC-Z/H.  CCD temperatures are relatively stable with an exception in 2013 when SSC-Z stopped working.}  \label{SSC_cooling}
\end{figure}

Fig.\,\ref{SSC_cooling} shows the thermal history of the radiators and CCDs.  The radiator-Z shows a large variation in day/night condition while the radiator-H shows a small variation.  LHP functions properly and the body of the SSC is cooled around -20$^\circ$C in orbit and is relatively stable.  Therefore, we select to keep the peltier current constant rather than the CCD temperature constant \cite{tsunemi2010}.  The actual temperature of the CCDs are around -65$^\circ$C that is well within the target temperature of -60$^\circ$C.

In 2013, we had one CPU malfunctioned, that controlled the SSC-Z.  It took about one year to recover.  During this period, the SSC-Z stopped working and became the same temperature to that of the SSC stand.  It is quite interesting to note that the temperature of the SSC-H increased in spite that the heat from the peltier devices became half.  This can be understood that the LHP efficiency degraded due to the decrease of the heat input.  After the recovery of the CPU, the thermal condition of the SSC also recovered as before.

\subsection{in orbit performance}
The SSC functions properly during the night-time and detects X-ray events and particle events.  However, it seriously suffers a signal overflows in the edge area of the CCD during the day time \cite{tsunemi2010}.  Sun light, particularly the infra-red, enters through the slit and is scattered inside the collimator.  Therefore, we focus on the night-time observation.  Furthermore, we can not obtain a useful data during the passage of the SAA and high background region.  The effective observation efficiency is about 30\%.

\begin{figure}
\sidecaption
   \includegraphics[width=2.5in]{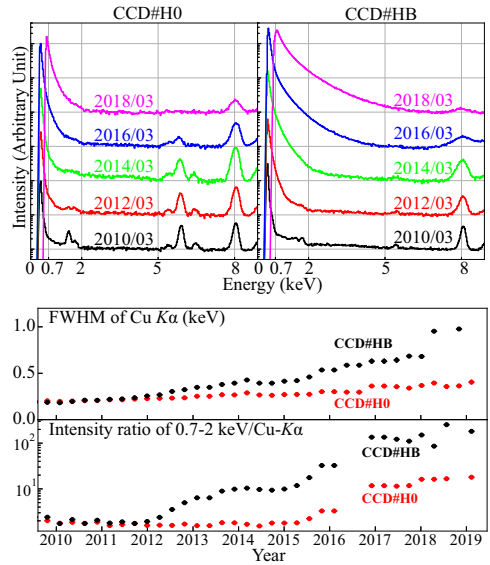}
\caption{Top panel shows the spectral evolution of CCD\#H0 and CCD\#HB.  We can see a degradation of the emission line features of Cu-K$\alpha$(8.09\,keV) and Mn-K$\alpha$(5.9\,keV) as well as the increase at low energy.  Bottom panel shows the degradation of the FWHM of Cu-K$\alpha$ and the intensity ratio between the low energy band (0.7-2\,keV) and Cu-K$\alpha$. }  \label{SSC_degradation}
\end{figure}

The increase of charge traps and dark current degrades SSC performance; a degradation of the energy resolution and an increase of the low energy background.  The energy resolution can be monitored by characteristic X-rays.  Radio-active sources of $^{55}$Fe are installed at the edge of the SSCU.  We also obtain continuous and uniform monitoring of Cu-K lines produced at the collimator by the incident particles.  The low energy background can be monitored by the X-ray spectrum.  Fig.\,\ref{SSC_degradation} shows the spectral history of the two CCDs; one (CCD\#HB) is at the center of the SSC and the other (CCD\#H0) is at the edge of the SSC.  The degradation of the CCD performance depends on the distance from the slit; CCD\#HB receives higher flux of charged particle than CCD\#H0.  The FWHM of Mn-K was 147\,eV and that of Cu-K was 170 eV at the time of launch \cite{tomida2011}.  A clear increase of the low energy is also noticed.

\subsection{SSC all sky map}
\paragraph{Point source catalog}
There are two types of instruments to perform all-sky survey.  One has a relatively narrow FOV; UHURU, Ariel\,V, HEAO\,I and ROSAT are of this type.  They can take a long time to cover the entire sky, up to half a year in the case of ROSAT, while individual sources are observed for a relatively short period.  As a result, they can miss sources that happen to be in quiet phase during its covering period.  The other has a much wider FOV, like 
MAXI.  It has a large fan beam FOV and scans the sky in every 92\,minutes.  Therefore, the source intensities listed in the MAXI catalog are accordingly the average ones over full survey period.  Hence, MAXI is far less likely to miss sources that fall in quiescence for a relatively short period than the other type of all-sky survey.

MAXI/SSC has much lower sensitivity than that of MAXI/GSC whereas it can cover the energy range below 2\,keV.  The SSC source catalog  \cite{tomida2016}, using the first 45-month data from 2010 August to 2014 April, contains two energy bands, 0.7-1.85\,keV (soft) and 1.85-7.0\,keV (hard).  The limiting sensitivity of 3 and 4\,mCrab are achieved and 140 and 138 sources are detected in the soft and hard energy bands, respectively.  Combining the two energy bands, 170 sources are listed in the MAXI/SSC catalog. They are 22 galaxies including AGNs, 29 cluster of galaxies, 21 supernova remnants, 75 X-ray binaries, 8 stars, 5 isolated pulsars, 9 non-categorized objects and 2 unidentified objects.  Comparing the soft-band sources with the ROSAT survey, which was performed about 20 years ago, 10\% of the cataloged sources are found to have changed the flux since the ROSAT era.

\begin{figure}[b]
\centering
    \includegraphics[width=2.2in]{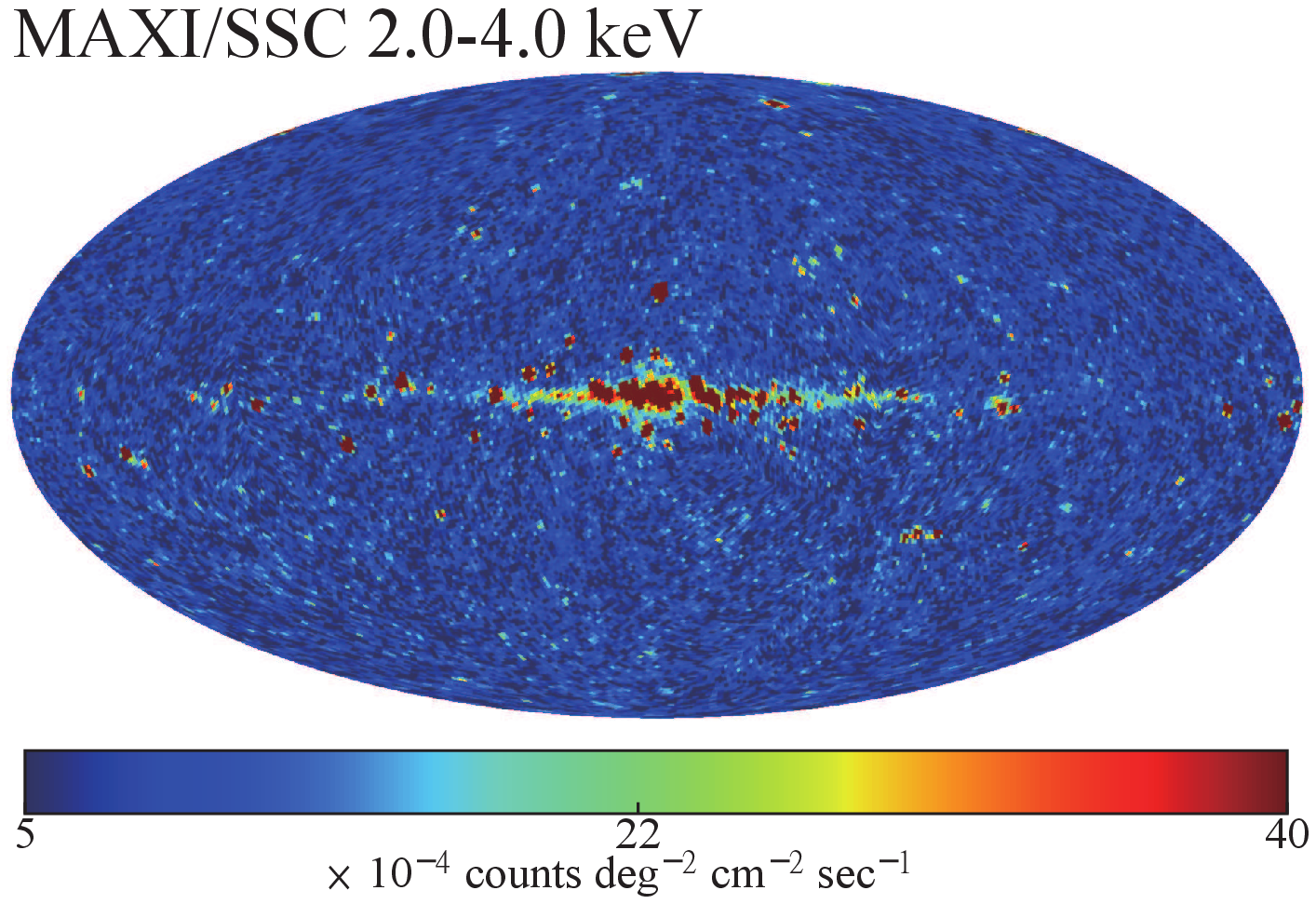}
    \includegraphics[width=2.2in]{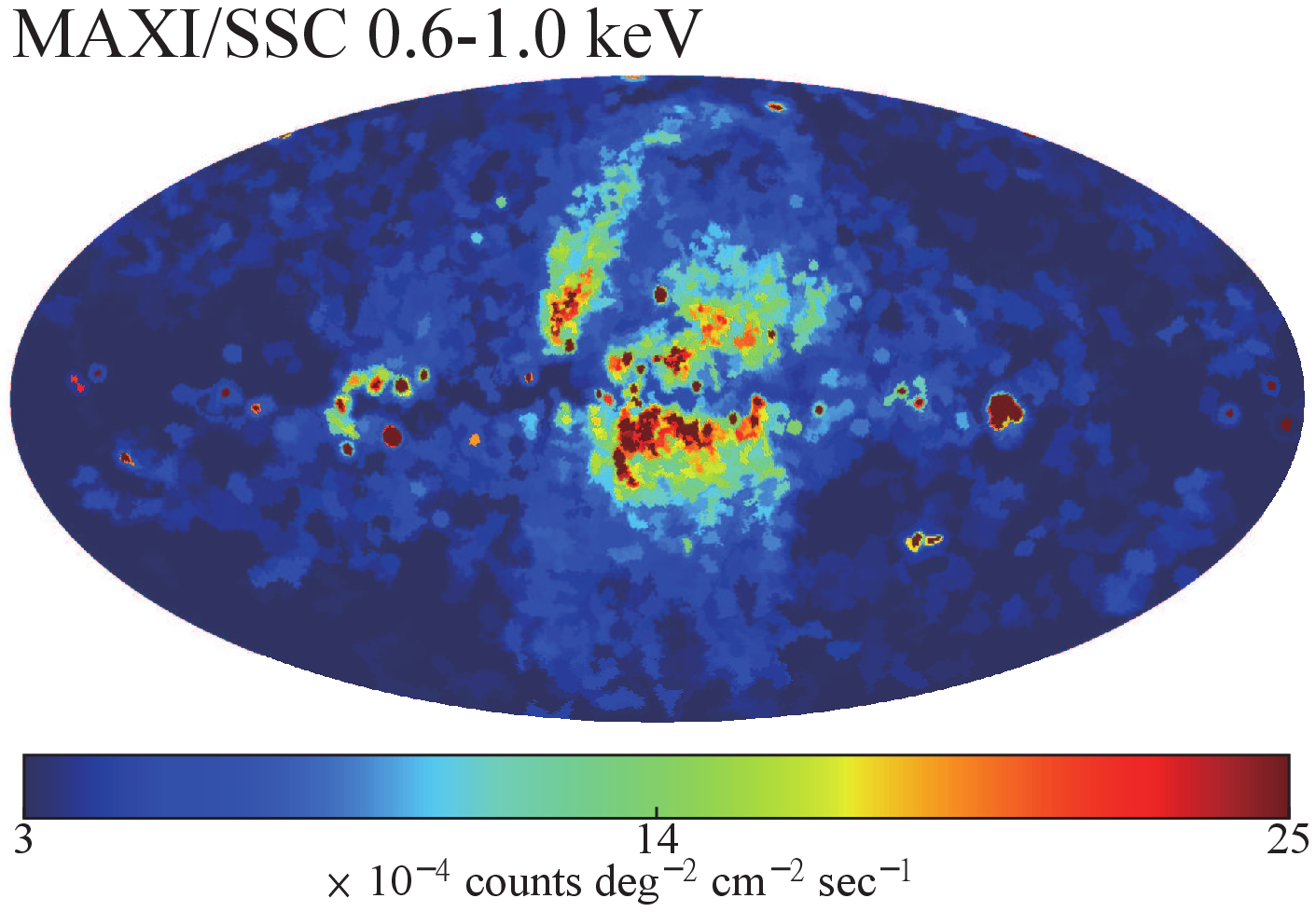}
\caption{ Left is the SSC all sky map in the X-ray energy bands of 2.0-4.0\,keV.  Bright point sources are saturated in this color map so that a uniform background can be seen.  Right is that in the X-ray energy bands of 0.7-1.0\,keV.  Various extended structures are seen.}
\label{SSC_band_map}
\end{figure}

\paragraph{Diffuse X-ray emission below 1\,keV}
SSC performed the all sky survey by using CCDs, particularly below 1\,keV.  After the background subtraction, we obtained the all sky maps in Fig.\,\ref{SSC_band_map}; the H-band (2-4\,keV) map and the L-band (0.7-1.0\,keV) map \cite{nakahira2020}.  There are many point sources in H-band map while they are saturated so that we can see a uniform background as well as a galactic ridge emission.  In the L-band map, we see many extended X-ray structures (EXSs); the Cygnus Super Bubble \cite{kimura2013}, the Vela SNR and a big one around the galactic center (GC).  The L-band map is very similar to that obtained by ROSAT about 20 years ago.

In the diffuse X-ray background, the Solar Wind Charge eXchange (SWCX) is believed to play a major part, particularly at low energy since it was first noticed from comet observations by ROSAT \cite{Cravens1997}.  It would affect the low energy (mainly below 1\,keV) X-ray data with many emission lines.  The intensity of the SWCX
must correlate with the solar activity while it is not yet completely confirmed.  The diffuse X-ray spectrum obtained by SSC can not be explained by the SWCX but by a thin thermal emission.  They are relatively stable intensity during the SSC observation.  The ROSAT observation was done near the peak of the solar cycle 22 while
that of the SSC was done near the minimum at the beginning of the solar cycle 24.  With taking into account these facts, we confirm that the diffuse X-ray emission at L-band is not generated by the SWCX but a celestial origin \cite{nakahira2020}.

A large EXS is seen in the direction of the GC.  There is some intensity inhomogeneity while they are thin thermal origin.  We only constrain that the interstellar absorption feature is at least 0.6 times that
of the galactic absorption feature.  The brighter part of EXS is well within a circle of about 50$^\circ$ radius whose center is about 25$^\circ$ away from the GC.  However, the distance to the EXS is not yet determined.  If
it is near the GC, a strong AGN activity is expected.  If it is far from the galactic center and near-by, a hypernova or a series of supernova is plausible.  A further observation is needed to understand its origin.

%\newpage
%\input{references}
%%\bibliographystyle{apj}

%\bibliography{reference}
%\bibliography{reference/reference, reference/tsunemi_NR_2014, reference/tsunemi_2014, reference/Satellite_CCD}

%%\bibliography{tsunemi_2014,Satellite_CCD}

%\end{document}

%3. MAXI data flow and novasearch 3p@@ª
\section{MAXI data flow and Nova Alert System} 
\subsection{Data Flow}

MAXI data are downloaded to NASA 
and transported to the JAXA Tsukuba Space Center (TKSC) in 5-6 seconds in total 
when the real-time contact is available \cite{negoro2016}.
About one-third of data are obtained during no real-time contact.
They are once stored in the onboard data recorder, the the High Rate Communication Outage Recorder (HCOR), in the ISS, and downloaded in the next contact one or two hours later.

\begin{table}{b}
\caption{Net 3-$\sigma$ detection limits$^a$ of the MAXI/GSC Nova-Alert System}
\label{tab:detlimit}       % Give a unique label
\begin{tabular}{p{2.4cm}p{2.2cm}p{2.2cm}p{2.2cm}p{2.2cm}}
\hline\noalign{\smallskip}
Time bins & 3--10 keV & 2--4 keV & 4--10 keV & 10--20 keV \\
\noalign{\smallskip}\svhline\noalign{\smallskip}
1 scan 			& 60--100  & 80--120 & 80--120 & 400$^b$\\
4 orbits ($\sim 6$ h) 	&  25--30 & 50--70 & 30--50 & 200\\
1 days  			& 12--15  & 25--30 & 15--20 & 100\\
4 days 			& 7--8 & 15--20 & 8--10 & 50\\
\noalign{\smallskip}\hline\noalign{\smallskip}
\end{tabular}
$^a$ The detection limits especially in shorter time bins depend on the number of detectors detected, incident angle, and background count rates. All the values are in unit of mCrab.
$^b$ This and below limits are rough estimation.
\end{table}

Normal GSC event data and house-keeping (HK) data are downloaded through the robust low-rate ($\sim$ 25 kbps) network with the Mil-1553B interface,
and mainly used for transient event search and health checks of the detectors.
%The minimum amount of data required are designed to download even if the available network is at the minimum speed of 25 kbps.
Complete GSC and SSC event data and HK data are transmitted
through the medium-rate network with the Ethernet interface whose maximum rate is 600 kbps.

The telemetry data are immediately processed at the TKSC, 
%resolved into each X-ray event and HK data, 
%Raw and processed GSC event data with celestial coordinates and a photon energy and
% the HK data are immediately % resolved from the telemetry data, and 
and stored into a \texttt{PostgreSQL} database in the MAXI photon-event database system (MAXI-DB) \cite{negoro2004}. 
Different from other all-sky X-ray monitors with coded masks, the slit-camera design enables us to 
determine the direction of each incident X-ray, and store the data photon by photon.
% and the data are delivered to the following systems.
% The MAXI-DB system provides X-ray event data  % the 1553B GSC and HK data 
% to the MAXI/GSC Nova-Alert system in real-time.
All the 1553B and Ethernet GSC, SSC, and HK data are provided from 
the ISAS/DARTS with the delay of about 1 hour and the NASA/HEASARC\footnote{
DARTS: https://www.darts.isas.jaxa.jp, HEASARC: https://heasarc.gsfc.nasa.gov}
for researchers to analyze the data.
GSC light curves of more than 400 objects and the on-demnad analysis tool \cite{nakahira2013}
are also available from the MAXI web site at RIKEN\footnote{http://maxi.riken.jp}.

\subsection{MAXI/GSC Nova-Alert System}

The MAXI/GSC Nova-Alert System \cite{negoro2016} is a system to find 
transient events promptly at the TKSC. The system consists of the {\it nova-search} system
to search significant time variable events and the {\it alert} system to select events sent 
from the nova-search system and to issues alerts.

\subsubsection{Nova-Search System}
The nova-search system divides the celestial sphere  into 49152 regions (pixels)
with the same surface area of $\sim0.8$ square degrees using the \texttt{HEALPix} C library 
\cite{gorski2005}. The system receives 1553B GSC event data every second from the MAXI-DB,
and accumulates the events into light curves at the corresponding celestial pixels.
The time-bins and the energy-bands of the light curves investigated are 
1-sec, 3-sec, 10-sec, 30-sec, 1-scan, 4-orbits, 1-day, and 4-days bins, 
and 3--10 keV,  2--4 keV,  4--10 keV, and 10--20 keV  bands, respectively.
Thus, the total number of the curves the system holds is $(49152$ pixels $\times\, 8$ 
time-scales $\times\, 4$ energy-bands $=) 1,572,864$.

Each light curve consists of ten time bins, nine of which  are used to calculate average count rates.
The rest of one and an additional one time bin are used to evaluate the latest count rates for real-time 
and delay-time (HCOR) data, respectively. The average count rate of the 1-scan bin data
is also used as those of the short time (1--30 sec) bin data. 
Every second, the nova-search system calculates net exposure time, effective area, and an expected background
count rate in each pixel where X-rays enter. 
Using those information, trigger thresholds are calculated in all the time-scale bins.
The criteria are given by chance probabilities typically less than or equal to $10^{-3 \sim -4}$ 
for the background (and source count) fluctuation. % to cause the observed number of counts.
Such low criteria result in a number of trigger events, $\sim 1$ events per second.

The nova-search system also play an important role as a real-time monitor developed by using 
the \texttt{GTK+} library.
The system gives us the real-time information about not only source activity but also detector health.
The system produces all-sky images every 1-orbit, 4-orbits, 1-d, and 4-d, 
which are utilized for duty scientists to confirm and/or find transient events (not automatically triggered) by eye.

%(or false alarm probability)
%%%%%
\subsubsection{Alert System}
%The alert system receives trigger information 
The alert system accumulates the trigger events at the sky-map regions (pixels) sent from the nova-search system.
The system issues alerts to the MAXI team (and the world for {\it burst} events) if
the system receives trigger events at the adjoining two pixels without the interruption for more than 12 ks (>2 orbits),
or the number of the triggers at one pixel containing a catalogued X-ray object exceeds 15 times.
The first condition comes from the fact that 
a bright point source is expected to split into two or more pixels because the PSF of the GSC with the full-width
at a half maximum of 1.5--2 deg is broader than the pixel size. 

The system determines the source position from the triggered-count-weighted pixel-center 
coordinates of HEALPix for the multi-pixel events, and from the pixel-center coordinates for the single-pixel events.
This and the positional insensitivity for the scan direction in the FOV within 1.5 degree make the position accuracy worse.
The resultant 90\% confidence error radius in the first E-mail alert is 0.7--0.8 arc-degree.
The final localization for the alert is determined by fitting a source image with the point spread function of GSC cameras, 
resulting in an uncertainty of 0.3--0.4 arc-degree for less variable transient sources.

The system classifies detected (multi- or single-pixel) events into {\it Burst, Alert, Warning}, and {\it Info} events 
before sending E-mail notifications according to the significance, trigger timescales, and identification.
For uncatalogued {\it Burst} events, the alert system automatically issues alerts to the world through the MAXI Mailing Lists
and {\it The Gamma-ray Coordinates Network} (GCN) in 10--30 sec after the onboard detection.
% Only uncatalogued {\it Burst} events are automatically circulated 
% through MAXI mailing lists and The Gamma-ray Coordinates Network (GCN).
 The duty scientists and senior staffs check all the events by eye, 
 % and manually evaluate the significance if necessary.
 % For transient events, the MAXI team manually 
and, if necessary, manually issue alerts in tens of minutes to a few hours after the detection.
The net 3-$\sigma$ detection limits for transient events with the Nova-Alert system are summarized in Table \ref{tab:detlimit}.

% If the alert system receives a triggered event at a sky-map pixel from the nova-search systems, the system records the DPTC time of the first received event as the trigger time. 
%The system also regards adjoining pixel events as one enhanced celestial source event and adds the events to the accumulated triggered-event 

%4. Scientific highlightsÚVµ¢¬Ê@3p@ª%
% 4.1 BHV¯  XTE 1752AÆ©AMAXI1535Æ©MAXI 1820 tanaka shibazakiÝ½¢ÈALCdËí¹Æ©B
% MAXI J0158
\section{Scientific Highlights}
The MAXI/GSC Nova-Alert system detected a number of transient events from short-term 
X/$\gamma$-ray burst events on time scales of seconds to long-term variabilities of X-ray binaries or AGNs on timescales
of hours to days \cite{negoro2016}. 
Some of them are soon reported to {\it The Astronomer's Telegram} (ATel) or GCN.
We also report to GCN upper-limits of gravitational wave events by LIGO-Virgo and high energy neutrino events 
by ICE-Cube notified through GCN.

\begin{figure}[t]
%  \begin{center}
  \includegraphics[width=115mm]{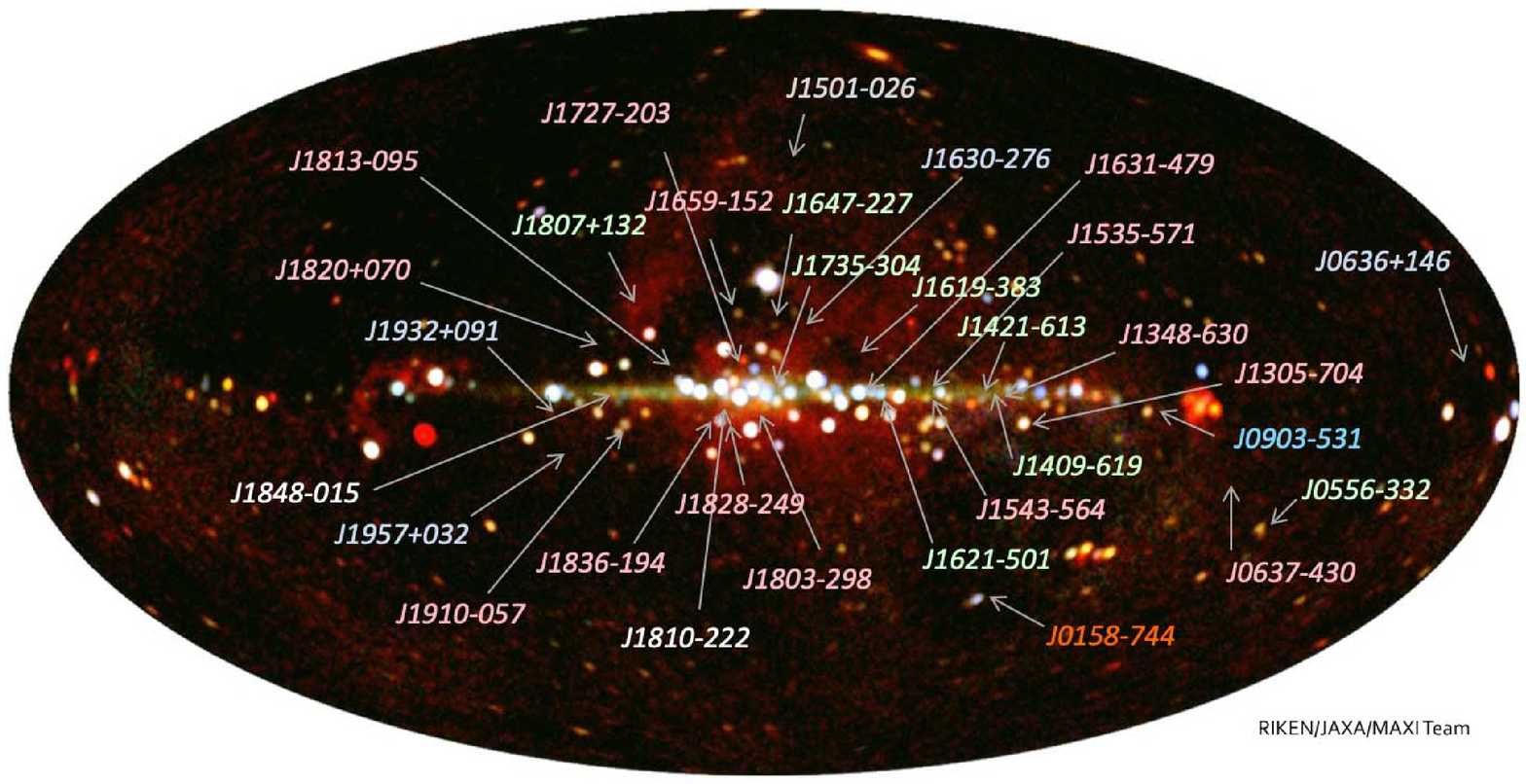} 
  \caption{31 X-ray novae MAXI newly discovered until the end of 2021. 
  Black holes are shown in pink labels, neutron stars in green, probable neutron stars in light blue, a pulsar in blue, 
  a white dwarf in red, and unknowns in white.
 The background image is the superposition of GSC and SSC false color images, obtained in the first 4.1 years and 3 years, respectively.
%\textcolor{black}{See figure 4 and 5 for the cross-section views.}
      }
  \label{fig:xnovaimg}
%  \end{center}
\end{figure}

\begin{figure}
%  \begin{center}
  \includegraphics[width=115mm]{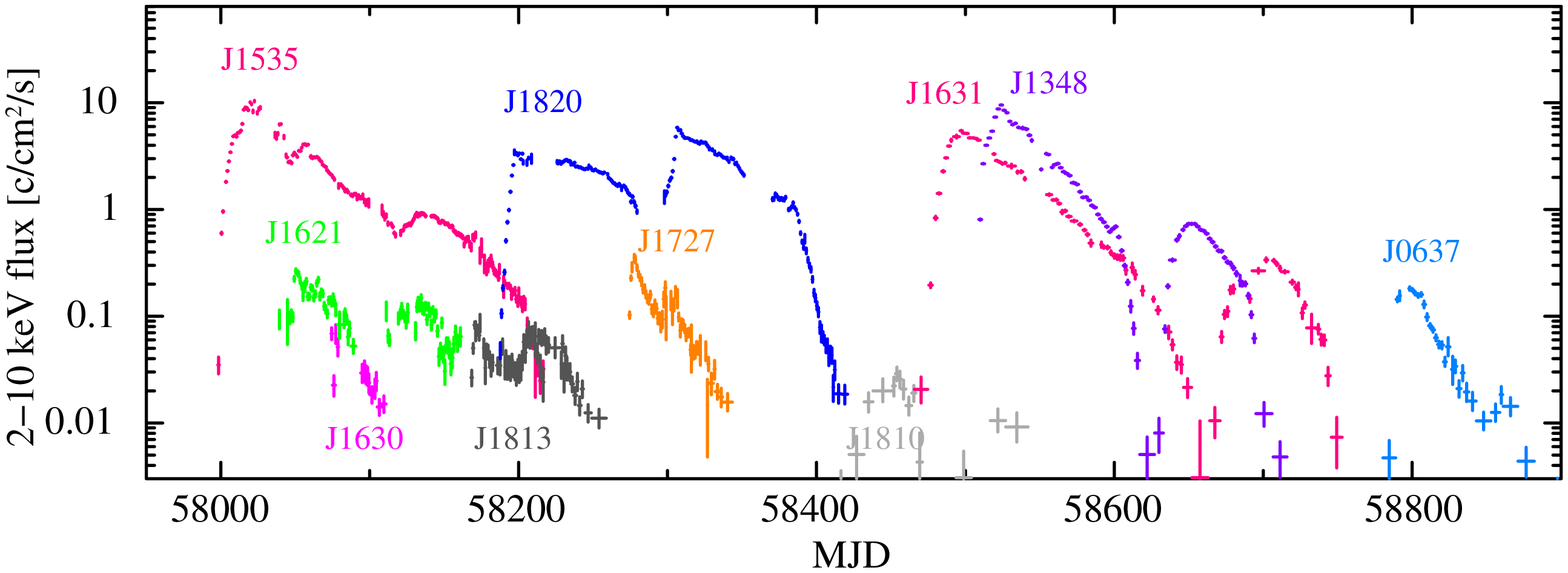} 
  \caption{2--10 keV light curves of  MAXI novae observed from 2017 April 7 (MJD 57850) to 2020 February 21 (MJD 58900). Some curves at low count rates are truncated to avoid the complexity due to the overlaps.
The count rate of the Crab nebula in the 2--10 keV band is 2.24 counts/cm$^2$/s.
%\textcolor{black}{See figure 4 and 5 for the cross-section views.}
      }
  \label{fig:xnovalc}
%  \end{center}
\end{figure}

\subsection{X-ray Bursts and Stellar Flares}

On average, one X-ray burst was detected every week, and a few super- or long-burst every year.
MAXI scanning observations every 92 min largely increased the number of the superbursts difficult to detect \cite{serino2016}.
One bright stellar flare mainly from multiple systems, RS CVn stars and dMe stars, is also detected every month \cite{tsuboi2016}.
MAXI observed the upper ends for stellar flares with the luminosity of 10$^{31-34}$ erg/s in the 2--20 keV band,
leading to the discovery of a universal correlation of $\tau \propto L_{\mathrm{X}}^{0.2}$ between the flare duration $\tau$ 
and the intrinsic X-ray luminosity $L_{\mathrm{X}}$ in the 0.1--100 keV band.

%Several large flares from AGNs also triggered the system every year.

\subsection{X-ray Novae and Short-lived Transients}

% Every year, MAXI newly discovered a few black hole or neutron star X-ray novae, and detected several known ones.
Until the end of 2021, MAXI newly discovered 31 X-ray novae or short-lived transients.
Figure \ref{fig:xnovaimg} displays positions of the new X-ray novae on the galactic coordinate;
14 black holes (candidates), 13 probable and confirmed neutron stars, 1 white dwarf, and 3 unknowns.
After the discovery, except a few novae, precise localizations were performed mostly by 
J. Kennea and his collaborators with Swift/XRT
in a few or several hours, sometimes, by NuSTAR or NICER. 

Figure \ref{fig:xnovalc} shows examples of long-term light curves of MAXI X-ray novae.
Continuous all-sky monitoring enables us to investigate detailed spectral evolution during the outbursts of these novae 
(e.g., \cite{nakahira1535} for J1535$-$571 and \cite{shidatsu1820} for J1820+070).
Peak fluxes of black hole X-ray novae range from several Crab to about 100 mCrab,
and the fluxes at soft-to-hard transitions are also different by more than one order of magnitude.
Taking account of the observational fact that the soft-to-hard transition occurs at a few percent of the Eddington luminosity \cite{motlagh2019}, 
these implies that MAXI sees most black hole X-ray novae in the whole our galaxy \cite{negoro2017}. 
% 2017 は maxi proc より場合よってはカット

% short-lived nova
MAXI also discovered short-lived transients lasting several hours or less, 
e.g., J1957+032 \cite{beri2019} and J0636-146. These are likely a kind of vary fast X-ray transient (VFXT). 
%Thanks to relatively high sensitivity to low energy band below 3 keV, 
Peculiar short-term soft X-ray transients only seen below 4 keV 
% with a thermal spectrum with a temperature of 500--700 eV 
were sometimes discovered.
One of the unpredictable discoveries is MAXI J0158$-$744 found in the out-skirt of the Small Magellanic Cloud, the first discovery of an unusually massive O--Ne white dwarf close to, or possibly over, the Chandrasekhar limit 
with a companion Be star \cite{morii2013}. 
MAXI Unidentified Short Soft X-ray Transients (MUSSTs) are those which were
detected only in one scan transit and no counterpart was found by prompt Swift/XRT followup observations. % \cite{musst}.
Such MUSST sources are not included to the above 31 new transients except for J1501$-$026 of highly 
significant detection. 
% GRB but soft? MUSST extra-galactic???
Recently one of the MUSSTs, GRB 140814A, is likely a merger-triggered core collapse supernova
\cite{dong2021}.

\begin{table}
\begin{center}
\caption{ The brightest X-ray transients above 1 Crab with MAXI in 2009-2021. In 2-20 keV in 1-day bin. They are transients backhole binaries or Be X-ray binaries.}\label{tab:top10}
\begin{tabular}{clcl}
\hline
  & name & $F_{\rm X}$ (Crab) & date \\
\hline
1& 4U 1543-475 &8.4 &2021/6/15 \\
2& Swift J0243.6+6124 &8.2 &2017/11/7 \\
3& MAXI J1535-571 &3.9 &2017/9/20 \\
4& A 0535+262     &2.9 &2020/11/23 \\
5& MAXI J1348-630 &2.6 &2019/2/9 \\
6& MAXI J1820+070 & 2.1 &2018/7/6 \\
7& MAXI J1631-479 & 1.3 &2019/1/8 \\
8& V 0332+53     & 1.0 &2015/7/27 \\
\hline
\end{tabular}
\end{center}
\end{table}

A few or several X-ray outbursts from known objects were also detected every year, e.g., V404 Cyg (GS 2023+338) and 
4U 1730$-$22 appearing after 26 and 49 years, respectively. The recurrence periods of X-ray novae are also an important
clue to know the number of compact stars in our Galaxy.
In Table \ref{tab:top10} the brightest X-ray transient with MAXI are listed.
This is in 2-20 keV and in 1-day bin. In the soft (2-4 keV) and hard (10-20 keV) bands, the rank changes. In the time-scale, the brightest persistent source is Sco X-1 (16 Crab) and the highest instant flux in the MAXI FOV was from SGR 1935+2154 (60 Crab in 0.2s 
on 2021/10/7).

\subsection{Extragalactic Transients and MAXI Catalogs}

\begin{figure*}[t]
%  \begin{center}
  \includegraphics[width=115mm]{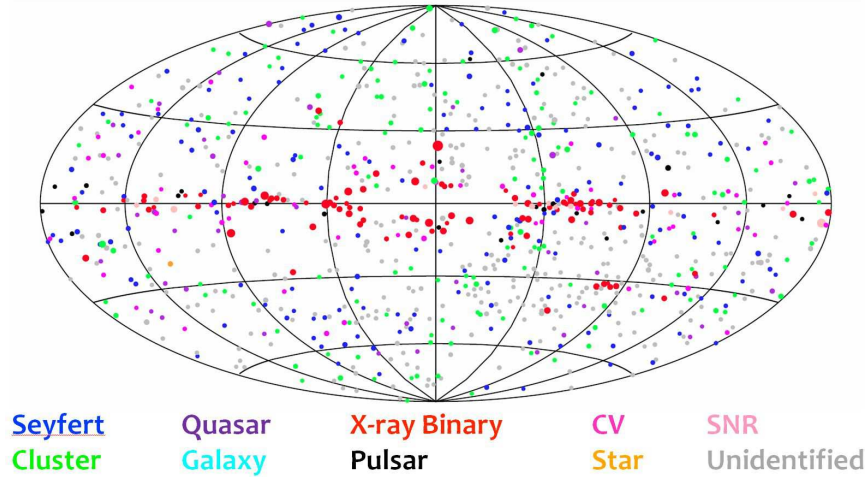} 
  \vspace{5mm}
  \caption{3MAXI catalog in the early 21st century. Total 896 sources.
  \cite{kawamuro2018}, \cite{hori2018}.}
  \label{fig:3maxi}
%  \end{center}
\end{figure*}

MAXI detected about one $\gamma$-ray burst every month. $\gamma$-ray bursts MAXI detected tend to have soft energy spectra, suggesting those bursts are off-axis events \cite{serino2014grb}. 
The second bright flare from the first realtime observed tidal disruption event of a star by a massive black hole 
Swift J164449.3+573451 \cite{burrows2011} was detected \cite{kimura2011}, which 
occurred about 34 hours after the first Swift/BAT detection. 
Longterm GSC monitoring observations of the whole sky allow us to detect 682 sources in the high Galactic latitude
with fluxes down to 0.48 mCrab ($\simeq 5.9 \times 10^{-12}$ erg\,cm$^{-2}$\,s$^{-1}$) in the 4--10 keV band
\cite{kawamuro2018}, and 221 sources in the low Galactic latitude with fluxes down to 0.43 mCrab \cite{hori2018}
in the first 7 years observations (2009/8 -- 2016/7), which are published as the 3MAXI catalog (Figure \ref{fig:3maxi}).
The flux level has almost reached the source confusion limit of GSC.
3MAXI catalogs are the new catalog in the 21st century, deepest in 4-10 keV band.
They also provide precious long-term light curves of variable sources.

%
%\begin{acknowledgement}
%acknowledgment text
%\end{acknowledgement}

%\section*{Appendix}
%\addcontentsline{toc}{section}{Appendix}
%
% appendix text

%\bibliographystyle{apj}
%\bibliography{mihara}

%\bibliography{tsunemi_2014,Satellite_CCD,mihara}

%\input{references}

\end{document}